\documentclass[12pt]{article}
\usepackage{amsfonts}
\usepackage{amssymb}
\usepackage{amsmath}
\usepackage[dvips]{graphicx}

\input xy
\xyoption{all}
\input{tcilatex}

\begin{document}

\begin{center}
{\Large \textbf{\ The WZW model }}

\bigskip

{\Large \textbf{\ on}}

\bigskip

{\Large \textbf{Random Regge Triangulations}}

\vspace{0.5 cm}

{\large \textsl{G.Arcioni}$^{a,}
$\footnote{{email : g.arcioni@phys.uu.nl}},} {\large \textsl{M. Carfora}$^{b,}
$\footnote{{email : mauro.carfora@pv.infn.it}},}

\vspace{0.2 cm}

{\large \textsl{C. Dappiaggi}$^{b,}$\footnote{{email :
claudio.dappiaggi@pv.infn.it}},} {\large \textsl{A. Marzuoli}$^{b,}$
\footnote{{email : annalisa.marzuoli@pv.infn.it}}.}

\vspace{0.5 cm}

$^{a}$Spinoza Institute and Institute for Theoretical Physics,\\[0pt]
Leuvenlaan 4 3584 CE Utrecht, The Netherlands

\vspace{0.1cm}

$^{b}$~Dipartimento di Fisica Nucleare e Teorica,

Universit\`{a} degli Studi di Pavia, \\[0pt]
and\\[0pt]
Istituto Nazionale di Fisica Nucleare, Sezione di Pavia, \\[0pt]
via A. Bassi 6, I-27100 Pavia, Italy

\vspace{0.1cm}

\bigskip
\end{center}

\begin{abstract}
By exploiting a correspondence between Random Regge triangulations (i.e.,
Regge triangulations with variable connectivity) and punctured Riemann
surfaces, we propose a possible characterization of the $SU(2)$ Wess-Zumino-Witten model
on a triangulated surface of genus $g$. Techniques of boundary CFT are used for the analysis of the quantum amplitudes of the model at level $\kappa=1$. These techniques provide a non-trivial algebra of boundary insertion operators governing a brane-like interaction between simplicial curvature and WZW fields. Through such a mechanism, we explicitly characterize the partition function of the model in terms of the metric geometry of the triangulation, and of the $6j$ symbols of the quantum group $SU(2)_Q$, at $Q=e^{\sqrt{-1}\pi /3}$. We briefly comment on the connection with bulk Chern-Simons theory. \end{abstract}



\newpage

\section{Introduction}

According to the holographic principle, in any theory combining quantum
mechanics with gravity the foundamental degrees of freedom are arranged in
such a way to give a quite peculiar upper bound to the total number of
independent quantum states. The latter are indeed supposed to grow
exponentially with the surface area rather than with the volume of the
system. The standard argument motivating such a view of the holographic
principle relies on the finitess of the black hole entropy: the number of
''bits'' of information that can be localized on the black hole horizon is
finite and determined by the area of the horizon. This led 't Hooft \cite
{'thooft} to conjecture the emergence of discrete structures describing the
degrees of freedom localized on the black hole horizon and an explicit and
significant example in the context of the S-matrix Ansatz program has been
given in \cite{'thooft2}. More recently \cite{'thooft3} the same author has
extended these considerations much beyond the physics of quantum black
holes, speculating that a sort of ''discrete'' quantum theory is at the
heart of the Planckian scale scenario, resembling a sort of cellular
automaton.

In view of these considerations, simplicial quantum gravity \cite{regge}
seems a rather natural framework within which discuss the holographic
principle. And, in this connection, some of us have recently proposed \cite
{arcioni} a holographic projection mechanism for a Ponzano-Regge model
living on a 3-manifold with non-empty fluctuating boundary. Related and very
interesting scenarios have been proposed also in \cite{various}. Although
such a discrete philosophy seems appealing, it must be said that \cite
{arcioni} fails short in bringing water to the mills of the holographic
principle since it is difficult to pinpoint the exact nature of the
(simplicial) boundary theory which holographically characterizes the bulk
Ponzano-Regge gravity. It is natural to conjecture that such a boundary
theory should be related with a $SU(2)$ WZW model, but the long-standing
problem of the lack of a suitable characterization of WZW models on metric
triangulated surfaces makes any such an identification difficult to carry
out explicitly.  As a matter of fact, quite
indipendently from any holographic issue, the formulation of  WZW theory
on a discretized manifold is a subject of considerable interest in itself,
and its potential field of applications is vast, ranging from the classical
connection with Chern-Simons theory and quantum groups, to moduli space
geometry and modern string theory dualities. It must be stressed that there
have been many attempts to characterize discrete WZW models starting from
discretized version of Chern-Simons theory (see e.g. \cite{kawamoto}). Rather than
providing yet another version of such a story,  here we do not start with
Chern-Simons theory and  work explicitly toward defining a procedure for
characterizing directly WZW models on triangulated surfaces. 

Many of the difficulties in blending WZW theory and Regge calculus (in any of its variants) 
stem from the usual technical problems in putting the
dynamics of $G$-valued fields ($G$ a compact Lie group) on a (randomly)
triangulated space: difficulties ranging from the correct simplicial
definitions of the domain of the $G$-fields, to their non-trivial dependence
from the topology of the underlying triangulation. A proper formulation
becomes much more feasible if one could introduce a description of the
geometry of randomly triangulated surface which is more analytic in spirit,
not relying exclusively on the minutiae of the combinatorics of simplicial
methods. Precisely with these latter motivation in mind some of us have
recently looked \cite{carfora}, \cite{carfora2} into the analytical aspects
of the geometry of (random) Regge triangulated surfaces. The resulting
theory turns out to be very rich and structured since it naturally maps
triangulated surfaces into pointed Riemann surfaces, and thus appears as a
suitable framework for providing a viable formalism for characterizing WZW
models on Regge (and dynamically) triangulated surfaces. 


The main goal of
this note is to apply the result of \cite{carfora2} to the introduction of  
$SU(2)$ WZW theory on metrically triangulated surfaces. In order to keep the
paper to a reasonable size and in order to coming quickly to grips with the
main points involved we limit ourselves here to the analysis of the
model in its non-trivial geometrical aspects, (some partial results in this connection have been announced in \cite{carfora3}), and to an explicit characterization of the partition function of the theory at level $\kappa=1$. Such a partition function has an interesting structure which directly involves the $6j$-symbols of the quantum group $SU(2)_Q$ at $Q=e^{\sqrt{-1}\pi /3}$, and depends in a non-trivial way from the metric geometry of the underlying triangulated surface. In its general features, it is not dissimilar from the (holographic) boundary partition function discussed in \cite{arcioni}, and owing to the explicit presence of the $SU(2)_Q$  $6j$-symbols one naturally expects for a rather direct  connection with a bulk Turaev-Viro model. Such a connection would frame in a nice combinatorial setup the known correspondence between the space of conformal blocks of the WZW model and the space of physical states of the bulk Chern-Simons theory. We do not reach such an objective here, nonetheless we pinpoint a few important elements which indicate that such a correspondence does indeed extend to our combinatorial framework. A detailed discussion of the relation with Chern-Simons theory, which puts to the fore the particular holographic issues that motivated us, will be presented elsewhere. 

Even if still incomplete in fulfilling its original holographic motivations, our analysis  of the WZW model on a triangulated surface exploits a few intermediate constructions and ideas that by themselves can be of intrinsic interest, since they put the whole subject in a wider perspective. In particular, the uniformization of a metric triangulated surface by means of a Riemann surface with (finite) cylindrical ends allows for an efficient use of boundary conformal field theory, and provides a rather direct connection with brane theory (here on group manifolds). We exploit such an interpretation for providing a description of the coupling mechanism between the (quantum) dynamics of the WZW fields and simplicial curvature. Roughly speaking, from the point of view of the dynamics of the WZW fields, (simplicial) curvature is seen as an exchange of closed strings between 2-branes in the group manifold. The interaction between the various closed string channels, (corresponding to the distinct curvature carrying vertices), is mediated by the operator product expansion between boundary insertion operators which are naturally associated with the metric ribbon graph defined by the 1-skeleton of the underlying triangulation. Note that, by uniformizing a random Regge triangulation with a (flat) Riemann surface with cylindrical ends, we are trading simplicial curvature for a modular parameter (the modulus of each cylindrical end turns out to be proportional to the conical angle of the corresponding vertex), and one is not plugging curvature by hands in the theory. Roughly speaking, gravity is indirectly read through the structure of the interaction between WZW fields and the modular parameters governing the closed string propagation between group branes. (Alternatively, by Cardy  duality, one can use an open string picture, with the cylindrical ends seen as closed loops diagrams of open strings with boundary points constrained to the group branes. In such a framework, the coupling with simplicial gravity can be seen as a Casimir like effect). These remarks suggest that simplicial methods have a role which is more foundational than usually assumed and that  they may provide a useful and reliable technique in a brane scenario.  
 \bigskip

Let us briefly summarize the content of the paper. In section 2, after
providing a few basic definitions, we recall the main results of \cite
{carfora} and \cite{carfora2} which feature prominently in the construction
of the WZW model on a Regge (and/or dynamical) triangulation. Here we introduce the correspondence between metric triangulated surfaces and the uniformization of a Riemann surface with cylindrical ends which is at the heart of the paper. 

In section 3
we discuss how we can naturally associate a $SU(2)$ WZW model to a (random)
Regge triangulation. The basic idea is to formulate WZW on the Riemann surface associated with the triangulation. In this way one can exploit all the known techniques of standard (i.e., continuum) WZW theory, and at the same time keep track of the relevant discrete aspects of the geometry of the original triangulation. A delicate point here concerns the imposition of suitable boundary conditions for the WZW fields at the cylindrical ends of the surface (the request for such boundary conditions cannot be avoided: it is a reflection of the fact that we cannot arbitrarily specify a WZW field at a conical vertex, there are monodromies to be respected). Our choice of boundary conditions is based on the remarkable analysis of
the boundary value theory of the WZW model due to K. Gaw\c{e}dzki \cite{gawedzki}. We discuss in detail all the steps needed for a proper characterization of the Zumino-Witten terms. As is known, this requires keeping track of the ambiguities in dealing with the extension of WZW maps to a three-dimensional bulk manifold bounded by the given Riemann surface. Such analysis naturally provides the proper set-up for moving to the quantum theory.

In section 4 we discuss the quantum amplitude of the model at level $\kappa =1$, (the reason for such a restriction are basically representation theoretic). By 
analysing  a natural factorization property of the WZW partition function on triangulated surface, we show how to exploit the results of \cite{gaberdiel} in order to characterize the  quantum amplitudes on each cylindrical end. We then discuss how such amplitudes interact along the ribbon graph associated with the underlying metrical triangulation. This step requires a rather detailed analysis of boundary insertion operators and of their operator product expansions along the vertices and edges of the ribbon graph. Here we are basically dealing with an application of  well-known sewing constraint techniques in boundary CFT, (relevant references for this part of the paper are \cite{lewellen},\cite{sagnotti},\cite{felder}). In particular, we exploit the connection between the OPE coefficient of such boundary operators and the $6j$-symbols of the quantum group $SU(2)_Q$, \cite{felder},\cite{gaume}. Finally, by factorizing a correlator of boundary insertion operators along the channels associated with the edge of the ribbon graph, we evaluate the partition function of the theory at level $\kappa=1$.
We conclude the paper with a a few remarks on the nature of such partition function indicating some of the  features which corroborate its natural connection with a (discretized) bulk Chern-Simons theory. 

\section{Uniformizing triangulated surfaces}

Let $M$ denote a closed 2-dimensional oriented manifold of genus $g$. A
(generalized) random Regge triangulation \cite{carfora} of $M$ is a
homeomorphism $|T_{l}|\rightarrow {M}$ where $T$ denote a $2$-dimensional
semi-simplicial complex with underlying polyhedron $|T|$ and where each edge 
$\sigma ^{1}(h,j)$ of $T$ is realized by a rectilinear simplex of variable
length $l(h,j)$. Note that since $T$ is semi-simplicial, the star of a vertex $
\sigma ^{0}(j)\in T$ (the union of all triangles of which $\sigma ^{0}(j)$
is a face) may contain just one triangle. Note also that the connectivity of 
$T$ is not a priori fixed as in the case of standard Regge triangulations
(see \cite{carfora} for details). In such a setting a (semi-simplicial)
dynamical triangulation $|T_{l=a}|\rightarrow {M}$ is a particular case \cite{ambjorn} of a
random Regge PL-manifold realized by rectilinear and equilateral simplices
of a fixed edge-length $l(h,j)=$ $a$, for all the $N_{1}(T)$ edges, where $
N_{i}(T)\in \mathbb{N}$ is the number of $i$-dimensional subsimplices $\sigma
^{i}(...)$ of $T$. Consider the (first) barycentric subdivision $T^{(1)}$ of $
|T_{l}|\rightarrow {M}$. The closed stars, in such a subdivision, of the
vertices of the original triangulation $|T_{l}|\rightarrow {M}$ form a
collection of $2$-cells $\{\rho ^{2}(i)\}_{i=1}^{N_{0}(T)}$ characterizing
the \emph{conical} Regge polytope $|P_{T_{l}}|\rightarrow {M}$ (and its
rigid equilateral specialization $|P_{T_{a}}|\rightarrow {M}$)
barycentrically dual to $|T_{l}|\rightarrow {M}$. The adjective conical
emphasizes that here we are considering a geometrical presentation $
|P_{T_{l}}|\rightarrow {M}$ of $P$ where the $2$-cells $\{\rho
^{2}(i)\}_{i=1}^{N_{0}(T)}$ retain the conical geometry induced on the
barycentric subdivision by the original metric structure of $
|T_{l}|\rightarrow {M}$. This latter is locally Euclidean everywhere except
at the vertices $\sigma ^{0}$, (the \textit{bones}), where the sum of the
dihedral angles, $\theta (\sigma ^{2})$, of the incident triangles $\sigma
^{2}$'s is in excess (negative curvature) or in defect (positive curvature)
with respect to the $2\pi $ flatness constraint. The corresponding deficit
angle $\varepsilon $ is defined by $\varepsilon =2\pi -\sum_{\sigma
^{2}}\theta (\sigma ^{2})$, where the summation is extended to all $2$
-dimensional simplices incident on the given bone $\sigma ^{0}$. In the case
of dynamical triangulations \cite{ambjorn} the deficit angles are generated
by the numbers $\#\{\sigma ^{2}(h,j,k)\bot \sigma ^{0}(k)\}$ of triangles
incident on the $N_{0}(T)$ vertices, the \textit{curvature assignments}, $
\{q(k)\}_{k=1}^{N_{0}(T)}\in \mathbb{N}^{N_{0}(T)}$, in terms of which we can
write $\varepsilon (k)=2\pi -\pi q(k)/3$.

It is worthwhile stressing that the natural automorphism group $Aut(P_{l})$
of $|P_{T_{l}}|\rightarrow {M}$, (\emph{i.e.}, the set of bijective maps
preserving the incidence relations defining the polytopal structure), is the
automorphism group of the edge refinement $\Gamma $ (see \cite{mulase}) of
the $1$-skeleton of the conical Regge polytope $|P_{T_{l}}|\rightarrow {M}$.
Such a $\Gamma$ is the $3$-valent graph 
\begin{equation}
\Gamma =\left( \{\rho ^{0}(h,j,k)\}\bigsqcup^{N_{1}(T)}\{W(h,j)\},
\{\rho^{1}(h,j)^{+}\}\bigsqcup^{N_{1}(T)}\{\rho ^{1}(h,j)^{-}\}\right) .
\label{uno}
\end{equation}
where the vertex set $\{\rho ^{0}(h,j,k)\}^{N_{2}(T)}$ is identified with
the barycenters of the triangles $\{\sigma ^{o}(h,j,k)\}^{N_{2}(T)}\in
|T_{l}|\rightarrow M$, whereas each edge $\rho ^{1}(h,j)\in \{\rho
^{1}(h,j)\}^{N_{1}(T)}$ is generated by two half-edges $\rho ^{1}(h,j)^{+}$
and $\rho ^{1}(h,j)^{-}$ joined through the barycenters $\{W(h,j)
\}^{N_{1}(T)}$ of the edges $\{\sigma ^{1}(h,j)\}$ belonging to the
original triangulation $|T_{l}|\rightarrow M$. The (counterclockwise)
orientation in the $2$-cells $\{\rho ^{2}(k)\}$ of $|P_{T_{l}}|\rightarrow {
M }$ gives rise to a cyclic ordering on the set of half-edges $\{\rho
^{1}(h,j)^{\pm }\}^{N_{1}(T)}$ incident on the vertices $\{\rho
^{0}(h,j,k)\}^{N_{2}(T)}$. According to these remarks, the
(edge-refinement of the) $1$-skeleton of $|P_{T_{l}}|\rightarrow {M}$ is a
ribbon (or fat) graph \cite{mulase}, \emph{viz.}, a graph $\Gamma $ together
with a cyclic ordering on the set of half-edges incident to each vertex of $
\Gamma$. Conversely, any ribbon graph $\Gamma $ characterizes an oriented
surface $M(\Gamma )$ with boundary possessing $\Gamma $ as a spine, (\emph{
i.e.}, the inclusion $\Gamma \hookrightarrow M(\Gamma )$ is a homotopy
equivalence). In this way (the edge-refinement of) the $1$-skeleton of a
generalized conical Regge polytope $|P_{T_{l}}|\rightarrow {M}$ is in a
one-to-one correspondence with trivalent metric ribbon graphs.

\begin{figure}[ht]
\begin{center}
\includegraphics[scale=.4]{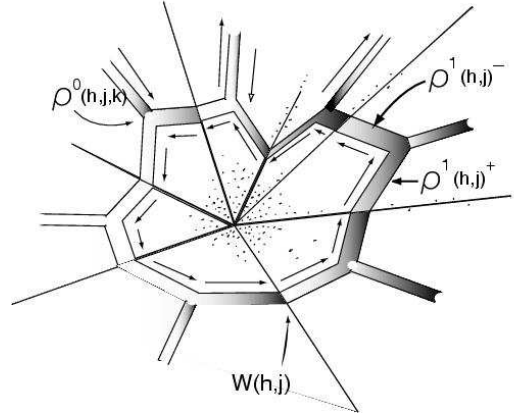}
\caption{The ribbon graph associated with the barycentrically dual polytope.}
\end{center}
\end{figure}

As we have shown in \cite{carfora}, \cite{carfora2} it is possible to naturally relax, (in the technical sense of the theory of geometrical structures \cite{Thurston}), the singular Euclidean structure associated with the conical  polytope $|P_{T_{l}}|\rightarrow {M}$ to a complex structure $((M;N_{0}),
\mathcal{C})$. Such a  relaxing is defined by exploiting \cite{mulase} the ribbon graph $\Gamma$ (see (\ref{uno})), and for
later use we need to recall some of the results of 
\cite{carfora2} by adopting a notation more suitable to our purposes.
Let  $\rho ^{2}(h)$, $\rho ^{2}(j)$, and $\rho ^{2}(k)$
respectively be the two-cells $\in |P_{T_{l}}|\rightarrow {M}$
barycentrically dual to the vertices  $\sigma ^{0}(h)$, $\sigma ^{0}(j)$,
and $\sigma ^{0}(k)$ of a triangle $\sigma ^{2}(h,j,k)\in |T_{l}|\rightarrow
M$. Let us denote by $\rho ^{1}(h,j)$ and $\rho ^{1}(j,h)$, respectively, the
oriented edges\ of $\rho ^{2}(h)$ and $\rho ^{2}(j)$ defined by 
\begin{equation}
\rho ^{1}(h,j)\bigsqcup \rho ^{1}(j,h)\doteq \partial \rho
^{2}(h)\bigcap_{\Gamma }\partial \rho ^{2}(j),
\end{equation}
\emph{i.e.}, the portion of the oriented boundary of \ $\Gamma $ intercepted
by the two adjacent oriented cells $\rho ^{2}(h)$ and $\rho ^{2}(j)$\ \
(thus $\rho ^{1}(h,j)\in \rho ^{2}(h)$ and $\ \rho ^{1}(j,h)\in \rho ^{2}(j)$
carry opposite orientations). Similarly, we shall denote by $\rho ^{0}(h,j,k)
$ the $3$-valent, cyclically ordered, vertex of $\Gamma $ defined by 
\begin{equation}
\rho ^{0}(h,j,k)\doteq \partial \rho ^{2}(h)\bigcap_{\Gamma }\partial \rho
^{2}(j)\bigcap_{\Gamma }\partial \rho ^{2}(k).
\end{equation}

\begin{figure}[ht]
\begin{center}
\includegraphics[scale=.4]{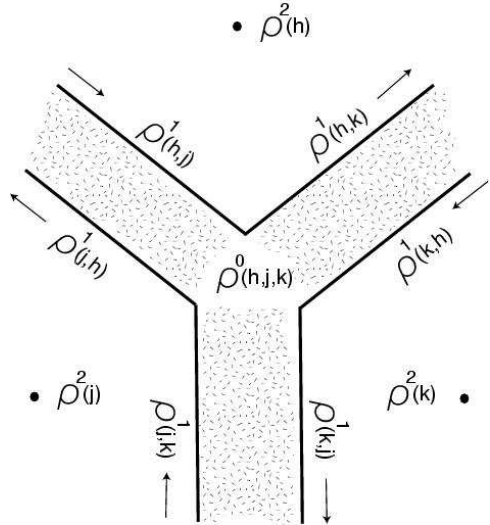}
\caption{The 2-cells,the oriented edges, and the oriented vertices of the conical dual polytope.}
\end{center}
\end{figure}

To the edge $\rho ^{1}(h,j)$ of $
\rho ^{2}(h)$ we associate \cite{mulase} a complex coordinate $
z(h,j)$ defined in the strip 
\begin{equation}
U_{\rho ^{1}(h,j)}\doteq \{z(h,j)\in \mathbb{C}|0<\func{Re}z(h,j)<L(h,j)\},
\label{nove}
\end{equation}
$L(h,j)$ being the length of the edge considered. The 
coordinate $w(h,j,k)$, corresponding to the $3$-valent vertex $\rho
^{0}(h,j,k)\in \rho ^{2}(h)$, is defined in the open set 
\begin{equation}
U_{\rho ^{0}(h,j,k)}\doteq \{w(h,j,k)\in \mathbb{C}|\;|w(h,j,k)|<\delta ,\;w(h,j,k)[\rho
^{0}(h,j,k)]=0\},
\end{equation}
where $\delta >0$ is a suitably small constant. Finally, the generic two-cell $\rho
^{2}(k)$ is parametrized in the unit disk 
\begin{equation}
U_{\rho ^{2}(k)}\doteq \{\zeta (k)\in \mathbb{C}|\;|\zeta (k)|<1,\;\zeta
(k)[\sigma ^{0}(k)]=0\},
\end{equation}
where $\sigma ^{0}(k)$ is the vertex $\in |T_{l}|\rightarrow M$
corresponding to the given two-cell.

\noindent We define the complex structure $((M;N_{0}),\mathcal{C})$ by
coherently gluing, along the pattern associated with the ribbon graph $
\Gamma $, the local coordinate neighborhoods $\{U_{\rho ^{0}(h,j,k)}\}_{(h,j,k)}^{N_{2}(T)}$, 
$\{U_{\rho ^{1}(h,j)}\}_{(h,j)}^{N_{1}(T)}$, and $\{U_{\rho
^{2}(k)}\}_{(k)}^{N_{0}(T)}$. Explicitly, (see \cite{mulase} for an elegant exposition of the general
theory and \cite{carfora}, \cite{carfora2} for the application to simplicial
gravity), let  $\{U_{\rho ^{1}(h,j)}\}$, $\{U_{\rho ^{1}(j,k)}\}$, $
\{U_{\rho ^{1}(k,h)}\}\ $ be the three generic open strips associated with
the three cyclically oriented edges $\rho ^{1}(h,j)$, $\rho ^{1}(j,k)$, $
\rho ^{1}(k,h)$ incident on the vertex $\rho ^{0}(h,j,k)$. Then the
corresponding coordinates $z(h,j)$, $z(j,k)$, and $z(k,h)
$\ are related to $w(h,j,k)$ by the transition functions 
\begin{equation}
w(h,j,k)=\left\{ 
\begin{tabular}{l}
$z(h,j)^{\frac{2}{3}},$ \\ 
$e^{\frac{2\pi }{3}\sqrt{-1}}z(j,k)^{\frac{2}{3}},$ \\ 
$e^{^{\frac{4\pi }{3}\sqrt{-1}}}z(k,h)^{\frac{2}{3}},$
\end{tabular}
\right. .  \label{glue1}
\end{equation}
Similarly, if $\{U_{\rho ^{1}(h,j_{\beta })}\}$, $\beta =1,2,...,q(k)$ are
the open strips associated with the $q(k)$ (oriented) edges $\{\rho
^{1}(h,j_{\beta })\}$ boundary of the generic polygonal cell $\rho ^{2}(h)$,
then the transition functions between the corresponding 
coordinate $\zeta (h)$ and the $\{z(h,j_{\beta })\}$ are given by \cite
{mulase} 
\begin{equation}
\zeta (h)=\exp \left( \frac{2\pi \sqrt{-1}}{L(h)}
\left( \sum_{\beta =1}^{\nu -1}L(h,j_{\beta })+z(h,j_{\nu })\right)
\right) ,\hspace{0.2in}\nu =1,...,q(h),  \label{glue2}
\end{equation}
with $\sum_{\beta =1}^{\nu -1}\cdot \doteq 0$, for $\nu =1$, and 
where $L(h)$ denotes the perimeter of $\partial (\rho ^{2}(h))$. By iterating such a construction for each vertex $\{\rho ^{0}(h,j,k)\}$ in the conical  polytope $|P_{T_{l}}|\rightarrow {M}$ we get a very explicit characterization of  $((M;N_{0}),\mathcal{C})$.

\begin{figure}[ht]
\begin{center}
\includegraphics[scale=.6]{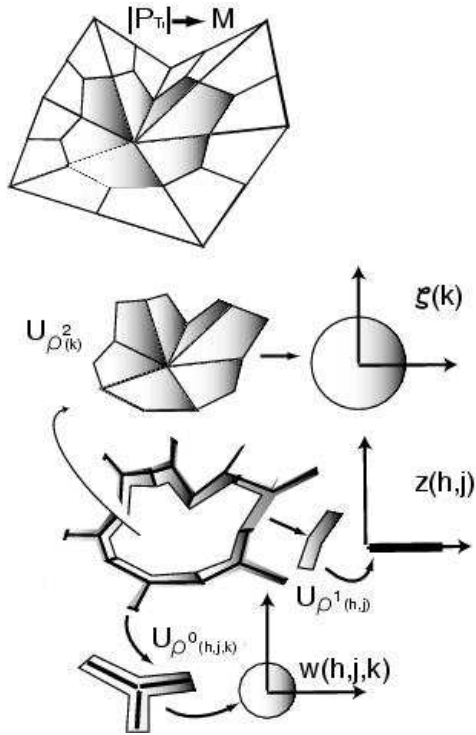}
\caption{The complex coordinate neighborhoods associated with the dual polytope.}
\end{center}
\end{figure}

Such a construction has a natural converse which allows us to describe the conical Regge polytope $|P_{T_{l}}|\rightarrow {M}$ as a uniformization of $((M;N_{0}),\mathcal{C})$. In this connection, the basic observation is that, in the complex coordinates introduced above, the ribbon graph $\Gamma $ naturally corresponds to 
a Jenkins-Strebel quadratic differential $\phi $
with a canonical local structure which is given by \cite{mulase} 
\begin{equation}
\phi \doteq \left\{ 
\begin{tabular}{l}
$\phi (h)|_{\rho ^{1}(h)}=dz(h)\otimes dz(h),$ \\ 
\\ 
$\phi (j)|_{\rho ^{0}(j)}=\frac{9}{4}w(j)dw(j)\otimes dw(j),$ \\ 
\\ 
$\phi (k)|_{\rho ^{2}(k)}=-\frac{\left[ L(k)\right]
^{2}}{4\pi ^{2}\zeta ^{2}(k)}d\zeta (k)\otimes d\zeta (k),$
\end{tabular}
\right.   \label{differ}
\end{equation}
where $L(k)$ denotes the perimeter of $\partial (\rho ^{2}(k))$, and  where $\rho ^{0}(h,j,k)$, $\rho ^{1}(h,j)$, $\rho ^{2}(k)$ run over the set of
vertices, edges, and $2$-cells of $|P_{l}|\rightarrow M$. If we denote by 
\begin{equation}
\Delta _{k}^{\ast }\doteq \{\zeta (k)\in \mathbb{C}|\;0<|\zeta (k)|<1\},
\label{puncdisk}
\end{equation}
the punctured disk $\Delta _{k}^{\ast}\subset U_{\rho ^{2}(k)}$, then for 
each given deficit angle $\varepsilon (k)=2\pi -\theta (k)$ we can introduce on each $\Delta _{k}^{\ast}$ the conical metric 
\begin{eqnarray}
ds_{(k)}^{2} &\doteq &\frac{\left[ L(k)\right] ^{2}}{
4\pi ^{2}}\left| \zeta (k)\right| ^{-2\left( \frac{\varepsilon (k)}{2\pi }
\right) }\left| d\zeta (k)\right| ^{2}=  \label{metrica} \\
&=& \left|
\zeta (k)\right| ^{2\left( \frac{\theta (k)}{2\pi }\right) }
|\phi (k)_{\rho ^{2}(k)}|,  \nonumber
\end{eqnarray}
where
\begin{equation}
|\phi (k)_{\rho ^{2}(k)}|=\frac{\left[ L(k)\right]
^{2} }{4\pi ^{2}|\zeta (k)|^{2}}|d\zeta (k)|^{2}.  \label{flmetr}
\end{equation}
is the standard cylindrical metric associated with the quadratic differential $\phi (k)_{\rho ^{2}(k)}$.

\begin{figure}[ht]
\begin{center}
\includegraphics[scale=.6]{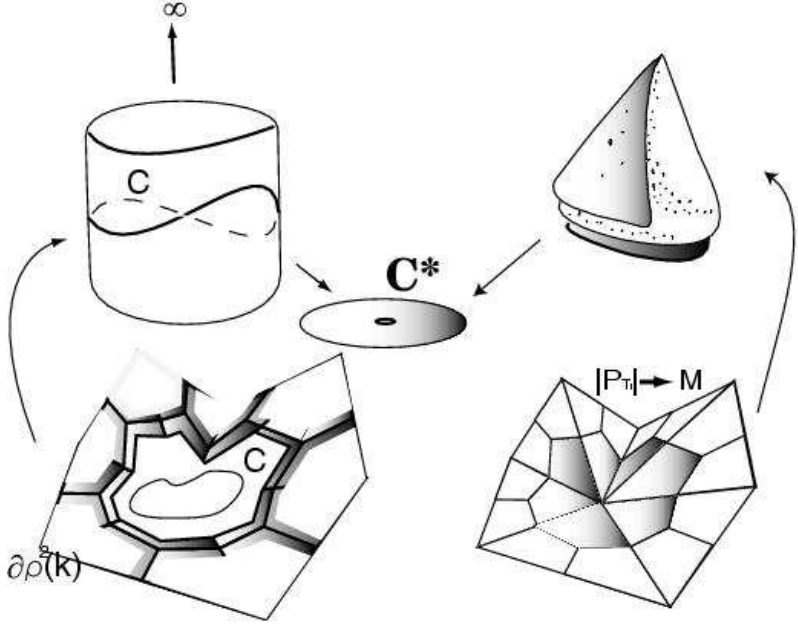}
\caption{The cylindrical and the conical metric over a polytopal cell.}
\end{center}
\end{figure}

In order to describe the geometry of the uniformization of 
$((M;N_{0}),\mathcal{C}))$ defined by $\{ds_{(k)}^{2}\}$,  let us consider the image in $((M;N_{0}),\mathcal{C}))$ of the generic triangle $\sigma ^{2}(h,j,k)\in |T_{l}|\rightarrow M$ of sides $\sigma ^{1}(h,j)$, $\sigma ^{1}(j,k)$, and $\sigma ^{1}(k,h)$. Similarly, let $W(h,j)$, $W(j,k)$, and $W(k,h)$ be the images of the respective barycenters, (see (\ref{uno})). Denote by 
$\widehat{L}(k)=|W(h,j)\rho^{0}(h,j,k)|$,  $\widehat{L}(h)=|W(j,k)\rho^{0}(h,j,k)|$, and
$\widehat{L}(j)=|W(k,h)\rho^{0}(h,j,k)|$, the lengths, in the metric $\{ds_{(k)}^{2}\}$, of the half-edges connecting the (image of the) vertex $\rho^{0}(h,j,k)$ of the ribbon graph $\Gamma$ with $W(h,j)$, $W(j,k)$, and $W(k,h)$. Likewise, let us denote by $l(\bullet,\bullet)$ the length of the corresponding side 
$\sigma ^{1}(\bullet,\bullet)$ of the triangle.   A direct computation involving the geometry of the medians of    
$\sigma ^{2}(h,j,k)$ provides
\begin{equation}
\begin{tabular}{ccc}
$\widehat{L}^{2}(j)$ & $=$ & $\frac{1}{18}l^{2}(j,k)+\frac{1}{18}
l^{2}(h,j)-\frac{1}{36}l^{2}(k,h )$ \\ 
$\widehat{L}^{2}(k)$ & $=$ & $\frac{1}{18}l^{2}(k,h)+\frac{1}{18}
l^{2}(j,k)-\frac{1}{36}l^{2}(h,j)$ \\ 
$\widehat{L}^{2}(h)$ & $=$ & $\frac{1}{18}l^{2}(h,j)+\frac{1}{18}
l^{2}(k,h)-\frac{1}{36}l^{2}(j,k)$ \\ 
&  &  \\ 
$l^{2}(k,h)$ & $=$ & $8\widehat{L}^{2}(h)+8\widehat{L}
^{2}(k)-4\widehat{L}^{2}(j)$ \\ 
$l^{2}(h,j)$ & $=$ & $8\widehat{L}^{2}(j)+8\widehat{L}
^{2}(h)-4\widehat{L}^{2}(k)$ \\ 
$l^{2}(j,k )$ & $=$ & $8\widehat{L}^{2}(k)+8\widehat{L}
^{2}(j)-4\widehat{L}^{2}(h)$
\end{tabular}
,
\end{equation}

\begin{figure}[ht]
\begin{center}
\includegraphics[scale=.5]{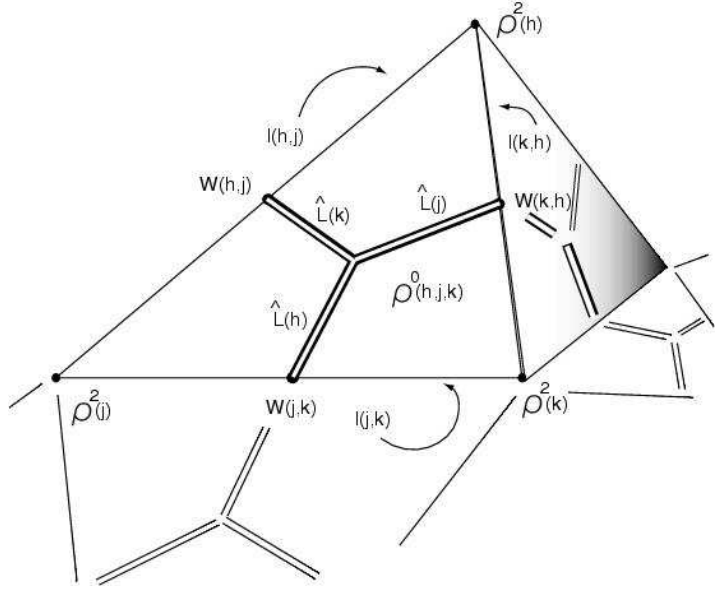}
\caption{The relation between the edge-lengths of the conical polytope and the edge-lenghts of the triangulation.}
\end{center}
\end{figure}

which allows to recover, as the indices $(h,j,k)$ vary, the metric geometry of  $|P_{T_{l}}|\rightarrow {M}$ and of its dual triangulation $|T_{l}|\rightarrow M$, from $((M;N_{0}),\mathcal{C});\{ds_{(k)}^{2}\})$.
 In this sense, the stiffening \cite{Thurston} of $((M;N_{0}),\mathcal{C})$ defined by the punctured Riemann surface 
\begin{gather}
((M;N_{0}),\mathcal{C});\{ds_{(k)}^{2}\})= \\
=\bigcup_{\{\rho ^{0}(h,j,k)\}}^{N_{2}(T)}U_{\rho ^{0}(h,j,k)}\bigcup_{\{\rho
^{1}(h,j)\}}^{N_{1}(T)}U_{\rho ^{1}(h,j)}\bigcup_{\{\rho
^{2}(k)\}}^{N_{0}(T)}(\Delta _{k}^{\ast },ds_{(k)}^{2}),  \nonumber
\end{gather}
is the uniformization of $((M;N_{0}),\mathcal{C})$ associated \cite{carfora2} with the conical Regge polytope $|P_{l}|\rightarrow M$.

\begin{figure}[ht]
\begin{center}
\includegraphics[scale=.5]{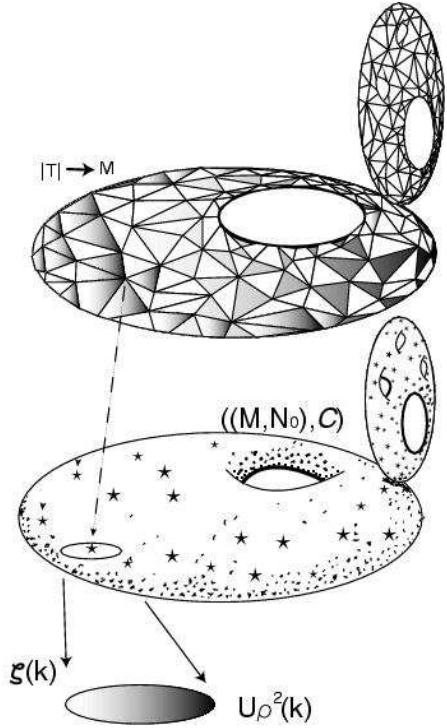}
\caption{The decorated punctured Riemann surface associated with a random Regge triangulation.}
\end{center}
\end{figure}

Although the correspondence between conical Regge polytopes and the above punctured Riemann surface is rather natural there is yet another uniformization representation of $|P_{l}|\rightarrow M$ which is of relevance in discussing conformal field theory on a given $|P_{l}|\rightarrow M$. The point is that the analysis of a CFT on a singular surface such as $|P_{l}|\rightarrow M$ calls for the imposition of suitable boundary conditions in order to take into account the conical singularities of the underlying Riemann surface $((M;N_{0}),\mathcal{C}, ds^2_{(k)})$. This is a rather delicate issue since conical metrics give rise to difficult technical problems in discussing the glueing properties of
the resulting conformal fields. In boundary conformal field theory, problems of this sort are taken care of (see e.g.[\cite{gawedzki}])  by (tacitly) assuming that a neighborhood of the possible boundaries is endowed with a cylindrical metric. In our setting such a prescription naturally calls into play the metric associated with the quadratic
differential $\phi $, and requires that we regularize into finite cylindrical ends the cones 
$(\Delta _{k}^{\ast },ds_{(k)}^{2})$.\ \  Such a  regularization is realized by noticing that if we introduce the annulus
\begin{equation}
\Delta _{\theta (k)}^{\ast }\doteq \left\{ \zeta (k)\in \mathbb{C}|e^{-\frac{
2\pi }{\theta (k)}}\leq |\zeta (k)|\leq 1\right\}\subset \overline{U_{\rho ^{2}(k)}},
\end{equation}
then the surface with boundary 
\begin{equation}
M_{\partial }\doteq ((M_{\partial };N_{0}),\mathcal{C})=\bigcup U_{\rho
^{0}(j)}\bigcup U_{\rho ^{1}(h)}\bigcup (\Delta _{\theta (k)}^{\ast },\phi
(k))
\end{equation}
defines the blowing up of the conical geometry of \ $((M;N_{0}),\mathcal{C}
,ds_{(k)}^{2})$ along the ribbon graph $\Gamma $.

\begin{figure}[ht]
\begin{center}
\includegraphics[scale=.5]{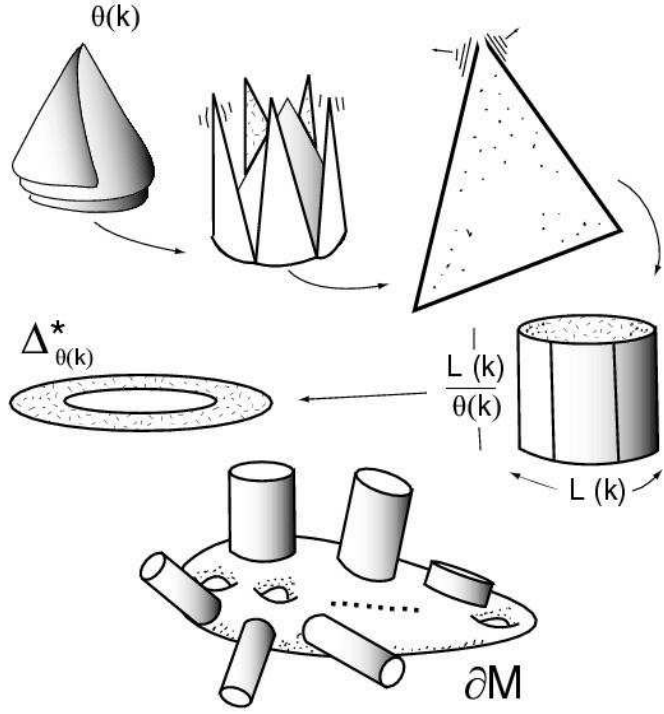}
\caption{Blowing up the conical geometry of the polytope into finite cylindrical ends generates a uniformized Riemann surface with cylindrical boundaries.}
\end{center}
\end{figure}

The metrical geometry of \ $(\Delta _{\theta (k)}^{\ast },\phi (k))
$ is that of a flat cylinder with a circumference of length given by $L(k)$ and heigth given by $L(k)/\theta (k)$, (this latter being the slant radius of the
generalized Euclidean cone $(\Delta _{k}^{\ast },ds_{(k)}^{2})$ of base circumference $L(k)$ and vertex
conical angle $\theta (k)$).We also have \ \ 
\begin{eqnarray}
\partial M_{\partial } &=&\bigsqcup_{k=1}^{N_{0}}S_{\theta (k)}^{(+)}, \\
\partial \Gamma  &=&\bigsqcup_{k=1}^{N_{0}}S_{\theta (k)}^{(-)}  \notag
\end{eqnarray}
where the circles 
\begin{eqnarray}
S_{\theta (k)}^{(+)} &\doteq &\left\{ \zeta (k)\in \mathbb{C}||\zeta
(k)|=e^{-\frac{2\pi }{\theta (k)}}\right\} , \\
S_{\theta (k)}^{(-)} &\doteq &\left\{ \zeta (k)\in \mathbb{C}||\zeta
(k)|=1\right\}   \notag
\end{eqnarray}
respectively denote the inner and the outer boundary of the annulus $\Delta
_{\theta (k)}^{\ast }$. 
Note that by collapsing $S_{\theta (k)}^{(+)}$ to a point we get back the original cones $(\Delta _{k}^{\ast },ds_{(k)}^{2})$.
 Thus, the surface with boundary $
M_{\partial }$ naturally corresponds to the ribbon graph $\Gamma $
associated with the 1-skeleton $K_{1}(|P_{T_{l}}|\rightarrow {M})$ of the
polytope $|P_{T_{l}}|\rightarrow {M}$, decorated with the finite
cylinders $\{\Delta _{\theta (k)}^{\ast },|\phi (k)|\}$. In such a
framework the conical angles $\{\theta (k)=2\pi -\varepsilon (k)\}$ appears
as (reciprocal of the) moduli $m_{k}$ of the annuli $\{\Delta _{\theta
(k)}^{\ast }\}$, 
\begin{equation}
m(k)=\frac{1}{2\pi }\ln \frac{1}{e^{-\frac{2\pi }{\theta (k)}}}=\frac{1}{\theta (k)}
\end{equation}
(recall that the modulus of an annulus $r_{0}<|\zeta |<r_{1}$ is defined by $\frac{1}{2\pi }\ln \frac{r_{1}}{r_{0}}$). According to these remarks we can equivalently represent the conical Regge polytope $|P_{T_{l}}|\rightarrow {M}$ with the uniformization  $((M;N_{0}),\mathcal{C});\{ds_{(k)}^{2}\})$ or with its blowed up version $M_{\partial }$. 

\section{The WZW model on a Regge polytope}

Let $G$ be a connected and simply connected Lie group. In order to make
things simpler we shall limit our discussion to the case $G=SU(2)$, this
being the case of more direct interest to us. Recall \cite{gawedzki} that
the complete action of the Wess-Zumino-Witten model on a closed Riemann
surface $M$ of genus $g$ is provided by 
\begin{equation}
S^{WZW}(h)=\frac{\kappa }{4\pi \sqrt{-1}}\int_{M}tr\left( h^{-1}\partial
h\right) \left( h^{-1}\overline{\partial }h\right) +S^{WZ}(h),
\end{equation}
where $h:M\rightarrow SU(2)$ denotes a $SU(2)$-valued field on $M$, $\kappa $
is a positive constant (the level of the model),  $tr(\cdot )$ is the
Killing form on the Lie algebra (normalized so that the root has length $
\sqrt{2}$) and $S^{WZ}(h)$ is the topological Wess-Zumino term needed \cite
{witten} in order to restore conformal invariance of the theory at the quantum level.
Explicitly, $S^{WZ}(h)$ can be characterized by extending the field $
h:M\rightarrow SU(2)$ to maps $\widetilde{h}:V_{M}\rightarrow SU(2)$ where $
V_{M}$ is a three-manifold with boundary such that $\partial V_{M}=M$, and
set 
\begin{equation}
S^{WZ}(h)=\frac{\kappa }{4\pi \sqrt{-1}}\int_{V_{M}}\widetilde{h}^{\ast
}\chi _{SU(2)},
\end{equation}
where $\widetilde{h}^{\ast }\chi _{SU(2)}$ denotes the pull-back to $V_{M}$
of the canonical 3-form on $SU(2)$

\begin{equation}
\chi _{SU(2)}\doteq \frac{1}{3}tr\left( h^{-1}dh\right) \wedge \left(
h^{-1}dh\right) \wedge \left( h^{-1}dh\right),  \label{treforma}
\end{equation}
(recall that for $SU(2)$, $\chi _{SU(2)}$ reduces to $4\mu _{S^{3}}$, where 
$\mu _{S^{3}}$ is the volume form on the unit 3-sphere $S^{3}$). As is well
known, $S^{WZ}(h)$ so defined depends on the extension $\widetilde{h}$ , the
ambiguity being parametrized by the period of the form $\chi _{SU(2)}$ over
the integer homology $H_{3}(SU(2))$. Demanding that the Feynman amplitude $
e^{-S^{WZW}(h)}$ is well defined requires that the level $\kappa$ is an
integer.

\begin{figure}[ht]
\begin{center}
\includegraphics[scale=.5]{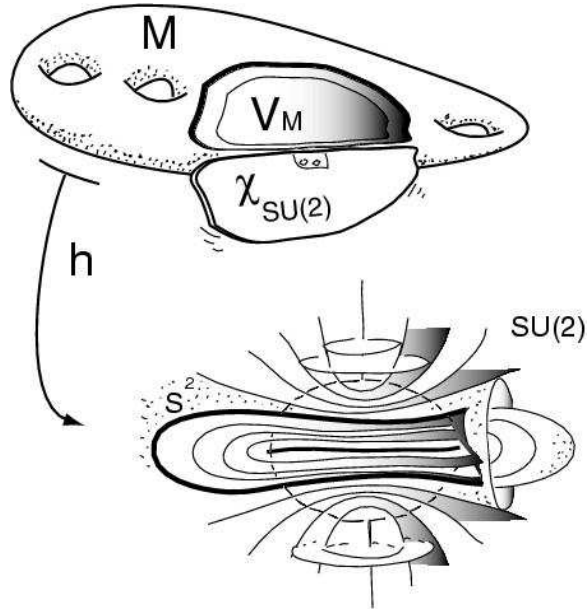}
\caption{The geometrical set up for the WZW model. The surface M opens up to show the associated handlebody. The group SU(2) is here shown as the 3-sphere foliated into (squashed) 2-spheres.}
\end{center}
\end{figure}

\subsection{Polytopes and the WZW model with boundaries}

From the results discussed in section 2, it follows that a natural strategy for introducing
the WZW model on the Regge polytope $|P_{T_{l}}|\rightarrow {\ M}$ is to
consider maps  $h:M_{\partial }\rightarrow SU(2)$ on the associated surface
with cylindrical boundaries $M_{\partial }\doteq ((M_{\partial };N_{0}),
\mathcal{C})$. Such maps $h$ should satisfy suitable boundary conditions 
on the (inner and outer) boundaries $\{S_{\theta (k)}^{(\pm )}\}$  of the annuli $\{\Delta
_{\theta (k)}^{\ast }\}$, corresponding to the (given) values of the $SU(2)$ field on the boundaries of the cells of $|P_{T_{l}}|\rightarrow {\ M}$ and on their barycenters, (the field being free to fluctuate in the cells). Among all
possible boundary conditions, there is a choice
which is particularly simple and which allows us to reduce the study of WZW
model on each given Regge polytopes to the (quantum) dynamics of WZW fields on the
finite cylinders (annuli) $\{\Delta _{\theta (k)}^{\ast }\}$ decorating the
ribbon graph $\Gamma $ and representing the conical cells of $|P_{T_{l}}|\rightarrow {\ M}$. 
Such an approach corresponds to first study the WZW model on $|P_{T_{l}}|\rightarrow {\ M}$ as a CFT. Its (quantum) states will then depend on the boundary conditions on the $SU(2)$ field $h$ on $\{S_{\theta (k)}^{(\pm )}\}$; roughly speaking such a procedure turns out to be equivalent to a prescription assigning an irreducible representation of $SU(2)$ to each barycenter of the given polytope  $|P_{T_{l}}|\rightarrow {\ M}$. Such representations are parametrized by the boundary conditions which, by consistency, turn out to be necessarily quantized. They are also parametrized by elements of the geometry of $|P_{T_{l}}|\rightarrow {\ M}$, in particular by the deficit angles. 
\bigskip

In order to carry over such a program, let us associate with each  inner boundary $S_{\theta
(i)}^{(+)}$ the $SU(2)$ Cartan generator 
\begin{equation}
\Lambda _{i}\doteq \frac{\lambda (i)}{\kappa }\mathbf{\sigma }_{3}\text{, with\ }\mathbf{
\sigma }_{3}=\left( 
\begin{array}{cc}
1 & 0 \\ 
0 & -1
\end{array}
\right) 
\end{equation}
where, for later convenience, $\lambda (i)\in \mathbb{R}$ has been normalized to the level $\kappa $, \
and let 
\begin{equation}
C_{i}^{(+)}\doteq \left\{ \gamma e^{2\pi \sqrt{-1}\Lambda _{i}}\gamma ^{-1}\;|\;\gamma
\in SU(2)\right\} .
\end{equation}
denote the (positively oriented) two-sphere $S^{2}_{\theta (i)}$ in $SU(2)$ representing the associated conjugacy class, (note that $C_{i}^{(+)}$ degenerates to a single point for the center of $SU(2)$). Such a prescription basically prevent out-flow of momentum across the boundary and has been suggested, in the framework of D-branes theory in \cite{alekseev}, (see also \cite{gawedzki}). Similarly, to the outer boundary $S_{\theta
(i)}^{(-)}$ we associate the conjugacy class $C_{i}^{(-)}=\overline{C_{i}^{(+)}}$ describing the conjugate two-sphere $\overline{S^{2}_{\theta (i)}}$  (with opposite orientation) in $SU(2)$ associated with  $S^{2}_{\theta (i)}$.   \ Given such data, 
we consider maps $h:M_{\partial
}\rightarrow SU(2)$ that satisfy the fully symmetric boundary conditions 
\cite{gawedzki2},
\begin{equation}
h(S_{\theta (i)}^{(\pm)})\subset C_{i}^{(\pm)}.
\label{achoice}
\end{equation}

\begin{figure}[ht]
\begin{center}
\includegraphics[scale=.4]{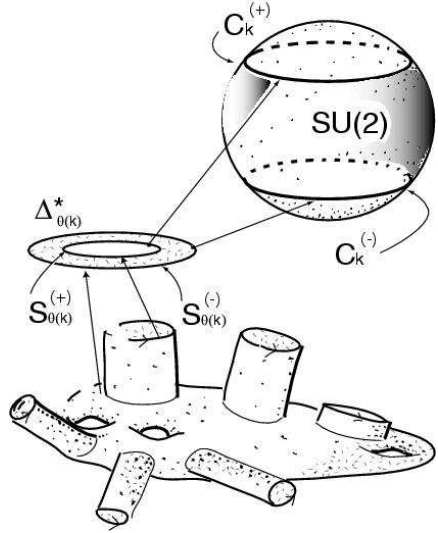}
\caption{The geometrical setup for SU(2) boundary conditions on each $(\Delta _{\theta (k)}^{\ast },\phi (k))$ decorating the 1-skeleton of $
|P_{T_{l}}|\rightarrow {M}$. For simplicity, the group SU(2) is incorrectly rendered; note that each circumference $C_k^{\pm}$ is actually a two-sphere, (or degenerates to a point).}
\end{center}
\end{figure}

Note that since $C_{i}^{(+)}$ and $C_{i}^{(-)}$ carry opposite orientations, the functions $h(S_{\theta (i)}^{(\pm)})$ are normalized to 
$h(S_{\theta (i)}^{(-)})h(S_{\theta (i)}^{(+)})={\mathbf e}$, (the identity $\in{SU(2)}$). The advantage of considering this subset of maps $h:M_{\partial
}\rightarrow SU(2)$ is that when
restricted to the boundary $\partial M_{\partial}$, ({\em i.e.}, to the inner conjugacy classes $C_{i}^{(+)}$), the 3-form $\chi _{SU(2)}$
(\ref{treforma}) becomes exact, and one can write 
\begin{equation}
\left. \chi _{SU(2)}\right| _{C_{i}}=d\omega _{i},
\end{equation}
where the 2-form $\omega _{i}$ is provided by 
\begin{equation}
\omega _{i}=tr(\gamma ^{-1}d\gamma )e^{2\pi \sqrt{-1}\Lambda _{i}}(\gamma
^{-1}d\gamma )e^{-2\pi \sqrt{-1}\Lambda _{i}}.
\end{equation}
In such a case, we can extend \cite{gawedzki} the map $h:M_{\partial
}\rightarrow SU(2)$ to a map ${\widehat{h}}:((M;N_{0}),\mathcal{C})\rightarrow SU(2)$ from the closed surface $((M;N_{0}),\mathcal{C})$ to $SU(2)$ in such a way that 
$\widehat{h}(\delta _{\theta (i)})\subset C_{i}^{(+)}$, where
\begin{equation}
\delta _{\theta (i)}\doteq \left\{ \zeta (i)\in \mathbb{C}|\;|\zeta (i)|\leq
e^{-\frac{2\pi }{\theta (i)}}\right\} 
\end{equation}
is the disk capping the cylindrical end $\{\Delta _{\theta (i)}^{\ast
},|\phi (i)|\}$, (thus $\partial \delta _{\theta (i)}=S_{\theta (k)}^{(+)}$ and 
$\Delta _{\theta
(i)}^{\ast }\cup \delta _{\theta (i)}\ \ \simeq \overline{U}_{\rho
^{2}(i)}$). In this connection note
that the boundary conditions $h(S_{\theta (i)}^{(+)})\subset C_{i}^{(+)}$
define elements of the loop group
\begin{equation}
\mathcal{L}_{(i)}SU(2)\doteq Map(S_{\theta (i)}^{(+)},SU(2))\simeq
Map(S^{1},SU(2)).
\end{equation}


\noindent Similarly, any other extension \ $h^{\prime }_{i}=\widehat{h}_{i}g$, ($g\in SU(2)$),
of $h$ over the capping disks $\delta _{\theta (i)}$, can be considered as an element of the group $Map(\delta
_{\theta (i)},SU(2))$. In the same vein, we can interpret $\widetilde{h}
_{i}=(\widehat{h}_{i},h_{i}^{\prime })$ as a map from the spherical double (see below) $
S_{i}^{2}$ of $\delta _{\theta (i)}$ into $SU(2)$, \emph{i.e.}, as an
element of the group $Map(S_{i}^{2},SU(2))$. It follows that each possible
extension of \ the boundary condition $h(S_{\theta (i)}^{(+)})$ fits into
the exact sequence of groups 
\begin{equation}
1\rightarrow Map(S_{i}^{2},SU(2))\rightarrow Map(\delta _{\theta
(i)},SU(2))\rightarrow Map(S_{\theta (i)}^{(+)},SU(2))\rightarrow 1.
\label{exseq}
\end{equation}
In order to discuss the properties of such  extensions we can proceed as follows, (see
\cite{gawedzki} for the analysis of these and
related issues in the general setting of boundary CFT).


Let us denote by $
V_{M}$,\ \ with $\partial V_{M}=((M,N_{0});\mathcal{C})$, the 3-dimensional
handlebody associated with the surface $((M,N_{0});\mathcal{C})$, and
corresponding to the mapping $\widehat{h}:((M,N_{0});\mathcal{C})\rightarrow
SU(2)\simeq S^{3}$ thought of as an \emph{immersion} in the 3-sphere. Since
the conjugacy classes $C_{i}^{(+)}$ are 2-spheres and the homotopy group $
\pi _{2}(SU(2))$ is trivial, we can further extend the maps $\widehat{h}$ to
a smooth function $\ \widehat{H}:V_{M}\rightarrow SU(2)$, (thus, by
construction \ $\widehat{H}(\delta _{\theta (i)})\subset $\ $C_{i}^{(+)}$).
Any such an extension can be used to pull-back to the handlebody $V_{M}$ the
Maurer-Cartan 3-form $\chi _{SU(2)}$ and it is natural to define the
Wess-Zumino term associated with $((M,N_{0});\mathcal{C})$ according to 

\begin{equation}
S_{|P_{T_{l}}|}^{WZ}(\widehat{h},\widehat{H})\doteq \frac{\kappa }{4\pi \sqrt{-1}}\int_{V_{M}}
\widehat{H}^{\ast }\chi _{SU(2)}-\frac{\kappa }{4\pi \sqrt{-1}}
\sum_{j=1}^{N_{0}}\int_{\delta _{\theta (j)}}\left. \widehat{h}\right|
_{\delta _{\theta (j)}}^{\ast }\omega _{j}.
\label{WZterm}
\end{equation}
In general, \ such a definition of \ $S_{|P_{T_{l}}|}^{WZ}(\widehat{h},\widehat{H})$\ depends
on the particular extensions $(\widehat{h},\widehat{H})$ we are considering,
and if we denote by $(h^{\prime }=\widehat{h}g,H^{\prime })$, $g\in SU(2)$,\
a different extension, then, by reversing the orientation of the handlebody $
V_{M}$ and of the capping disks $\delta _{\theta (j)}$ over which $
S_{|P_{T_{l}}|}^{WZ}(h^{\prime },H^{\prime })$ is
evaluated, the difference between the resulting WZ terms can be written as 
\begin{gather}
 S_{|P_{T_{l}}|}^{WZ}(\widehat{h},\widehat{H}
)-S_{|P_{T_{l}}|}^{WZ}(h^{\prime },H^{\prime })=
\label{diff} \\
=\frac{\kappa }{4\pi \sqrt{-1}}\left( \int_{V_{M}}\widehat{H}^{\ast }\chi
_{SU(2)}+\int_{V_{M}^{(-)}}H^{\prime \ast }\chi _{SU(2)}\right) -  \notag \\
-\frac{\kappa }{4\pi \sqrt{-1}}\sum_{j=1}^{N_{0}}\left( \int_{\delta
_{\theta (j)}}\left. \widehat{h}\right| _{\delta _{\theta (j)}}^{\ast
}\omega _{j}+\int_{\delta _{\theta (j)}^{(-)}}\left. h^{\prime }\right|
_{\delta _{\theta (j)}}^{\ast }\omega _{j}\right) .  \notag
\end{gather}
Note that 
\begin{equation}
(V_{M},\widehat{H})\cup (V_{M}^{(-)},H^{\prime })=(\widetilde{V}_{M},
\widetilde{H})
\end{equation}
is the 3-manifold (ribbon graph) double of $V_{M}$ \ endowed with the
extension $\widetilde{H}\doteq (\widehat{H},H^{\prime })$ \ \ and 
\begin{equation}
(\delta _{\theta (j)},\widehat{h_{j}})\cup (\delta _{\theta
(j)}^{(-)},h_{j}^{\prime })=(S_{j}^{2},\widetilde{h_{j}}),
\end{equation}
are the 2-spheres defined by doubling the capping disks $\delta _{\theta (j)}
$, decorated with the extension $\widetilde{h}_{j}\doteq (\widehat{h}
_{j},h_{j}^{\prime })\in C_{j}^{(+)}$. By construction $(\widetilde{V}_{M},
\widetilde{H})$ is such that $\partial (\widetilde{V}_{M},\widetilde{H}
)=\cup _{j=1}^{N_{0}}(S_{j}^{2},\widetilde{h_{j}})$ so that we can
equivalently write (\ref{diff}) as
\begin{gather}
 S_{|P_{T_{l}}|}^{WZ}(\widehat{h},\widehat{H}
)-S_{|P_{T_{l}}|}^{WZ}(h^{\prime },H^{\prime })=
\frac{\kappa }{4\pi \sqrt{-1}}\int_{\widetilde{V}_{M}}\widetilde{H}^{\ast
}\chi _{SU(2)}-  \label{diff2} \\
-\frac{\kappa }{4\pi \sqrt{-1}}\sum_{j=1}^{N_{0}}\int_{S_{j}^{2}}\widetilde{h
}^{\ast }\omega _{j}.  \notag
\end{gather}
To such an expression we add and subtract 
\begin{equation}
\frac{\kappa }{4\pi \sqrt{-1}}\sum_{j=1}^{N_{0}}\int_{B_{j}^{3}}\widetilde{
H_{j}}^{\ast }\chi _{SU(2)}
\end{equation}
where $B_{j}^{3}$ are 3-balls such that $\partial B_{j}^{3}=S_{j}^{2(-)}$,
(the boundary orientation is inverted so that we can glue such $B_{j}^{3}$
to the corresponding boundary components of $\widetilde{V}_{M}$), and $
\widetilde{H_{j}}$ are corresponding extensions of $\widetilde{H}$ with $
\widetilde{H_{j}}|_{S_{j}^{2}}=\widetilde{h}_{j}$. Since $\widetilde{V}
_{M}\cup $\ $B_{j}^{3}$ results in a closed 3-manifold $W^{3}$, we
eventually get 
\begin{gather}
S_{|P_{T_{l}}|}^{WZ}(\widehat{h},\widehat{H}
)-S_{|P_{T_{l}}|}^{WZ}(\eta )(h^{\prime },H^{\prime })=
\frac{\kappa }{4\pi \sqrt{-1}}\int_{W^{3}}\widetilde{H}^{\ast }\chi _{SU(2)}-
\\
-\frac{\kappa }{4\pi \sqrt{-1}}\sum_{j=1}^{N_{0}}\left( \int_{B_{j}^{3}}
\widetilde{H_{j}}^{\ast }\chi _{SU(2)}-\int_{\partial B_{j}^{3}}\widetilde{h}
^{\ast }\omega _{j}\right) ,  \notag
\end{gather}
where we have rewritten the integrals over $S_{j}^{2}$\ appearing in (\ref
{diff2}) as integrals over $\partial B_{j}^{3}=S_{j}^{2(-)}$, (hence the
sign-change). \ This latter expression shows that inequivalent extensions
are parametrized by the periods of $(\chi _{SU(2)},\omega _{j})$ over the
relative integer homology groups $H_{3}(SU(2),\cup _{j=1}^{N_{0}}C_{j})$.
Explicitly, the first term provides 
\begin{equation}
\frac{\kappa }{4\pi \sqrt{-1}}\int_{W^{3}}\widetilde{H}^{\ast }\chi _{SU(2)}=
\frac{\kappa }{4\pi \sqrt{-1}}\int_{\widetilde{H}(W^{3})}\chi _{SU(2)}=\frac{
\kappa }{4\pi \sqrt{-1}}\int_{S^{3}}\chi _{SU(2)}.
\end{equation}
Since $\int_{S^{3}}\chi _{SU(2)}=8\pi ^{2}$, we get $\frac{\kappa }{4\pi 
\sqrt{-1}}\int_{W^{3}}\widetilde{H}^{\ast }\chi _{SU(2)}=-2\pi \kappa \sqrt{-1}$
. Each addend in the second group of terms yields 
\begin{gather}
\frac{\kappa }{4\pi \sqrt{-1}}\left( \int_{B_{j}^{3}}\widetilde{H_{j}}^{\ast
}\chi _{SU(2)}-\int_{\partial B_{j}^{3}}\widetilde{h}^{\ast }\omega
_{j}\right) =  \label{remterm} \\
=\frac{\kappa }{4\pi \sqrt{-1}}\left( \int_{\widetilde{H}_{j}(B_{j}^{3})}
\chi _{SU(2)}-\int_{\widetilde{h}(\partial B_{j}^{3})}\omega _{j}\right) . 
\notag
\end{gather}


The domain of integration $\widetilde{h}(\partial B_{j}^{3})$ is the
2-sphere \ $C_{j}\subset SU(2)$ associated with the given conjugacy class,
whereas $\widetilde{H}_{j}(B_{j}^{3})$ is one of the two 3-dimensional balls
in $SU(2)$ with boundary $C_{j}$. In the defining representation of $
SU(2)\doteq \{x_{0}\mathbf{I}+\sqrt{-1}\sum x_{k}\mathbf{\sigma }_{k}|\;x_{0}^{2}+\sum
x_{k}^{2}=1\}$, the conjugacy classes $C_{j}$ are defined by $x_{0}=\cos 
\frac{2\pi \lambda (j)}{\kappa }$ with $0\leq \frac{2\pi \lambda (j)}{\kappa 
}\leq \pi $, whereas the two 3-balls $\widetilde{H}_{j}(B_{j}^{3})$ bounded
by $C_{j}$ are defined by $x_{0}\geq \cos \frac{2\pi \lambda (j)}{\kappa }$
and $x_{0}\leq \cos \frac{2\pi \lambda (j)}{\kappa }$. An explicit
computation \cite{gawedzki} over the ball $x_{0}\geq \cos \frac{2\pi \lambda (j)}{\kappa }$
shows that (\ref{remterm}) is provided by $-4\pi \lambda (j)\sqrt{-1}$, and
by $4\pi \sqrt{-1}(\frac{\kappa }{2}-\lambda (j))$ for $x_{0}\leq \cos \frac{
2\pi \lambda (j)}{\kappa }$, \ respectively. From these remarks it follows
that 
\begin{equation}
S_{|P_{T_{l}}|}^{WZ}(\widehat{h},\widehat{H}
)-S_{|P_{T_{l}}|}^{WZ}(h^{\prime },H^{\prime })\in
2\pi \sqrt{-1}\mathbb{Z}
\end{equation}
as long as $\kappa $ is an integer, and $0\leq \lambda (j)\leq \frac{\kappa 
}{2}$ with $\lambda (j)$ integer or half-integer; in such a case the
exponential of \ the WZ term $S_{|P_{T_{l}}|}^{WZ}(\widehat{h},
\widehat{H})$ is independent from the chosen extensions $(\widehat{h},
\widehat{H})$, and we can unambiguosly write $S_{|P_{T_{l}}|}^{WZ}(\widehat{h})$.
\bigskip
It follows from such remarks that we can define the $SU(2)$ WZW action on 
$|P_{T_{l}}|\rightarrow {M}$ according to
\begin{equation}
S_{|P_{T_{l}}|}^{WZW}(\widehat{h})\doteq \frac{\kappa }{4\pi \sqrt{-1}}
\int_{((M;N_{0}),\mathcal{C})}tr\left( \widehat{h}^{-1}\partial \widehat{h}
\right) \left( \widehat{h}^{-1}\overline{\partial }\widehat{h}\right)
+S_{|P_{T_{l}}|}^{WZ}(\widehat{h}).  \label{WZWact}
\end{equation}
where the WZ term $S_{|P_{T_{l}}|}^{WZ}(\eta )$ is provided by (\ref{WZterm}). It is worthwhile stressing that the condition $0\leq \lambda (j)\leq \frac{\kappa}{2}$ plays here the role of a quantization condition on the possible set of boundary conditions allowable for the WZW model on $|P_{T_{l}}|\rightarrow {M}$. Qualitatively, the situation is quite similar to the dynamics of branes on group manifolds, where in order to have stable, non point-like branes, we need a non vanishing $B$-field generating a NSNS 3-form $H$, (see e.g. \cite{schomerus}), here provided by $\omega_{j}$ and $\chi_{SU(2)}$, respectively. In such a setting, stable branes on $SU(2)$ are either point-like (corresponding to elements in the center $\pm e$ of $SU(2)$), or 2-spheres associated with a discrete set of radii. In our approach, such branes appear as the geometrical loci describing boundary conditions for WZW fields evolving on singular Euclidean surfaces. It is easy to understand the connection between the two formalism: in our description of the $\kappa$-level $SU(2)$ WZW model on $|P_{T_{l}}|\rightarrow {M}$ we can interpret the $SU(2)$ field as parametrizing an immersion of $|P_{T_{l}}|\rightarrow {M}$ in $S^{3}$ (of radius $\simeq \sqrt{\kappa}$). In particular, the annuli $\Delta^{*}_{\theta(i)}$ associated with the  ribbon graph boundaries $\{\partial\Gamma_i\}$ can be thought of as sweeping out in $S^{3}$ closed strings which couples with the branes defined by  $SU(2)$ conjugacy classes.

\section{The Quantum Amplitudes at $\kappa=1$}

We are now ready to discuss the quantum properties of the fields $\widehat{h}$ involved in the above characterization of the $SU(2)$ WZW action on $|P_{T_{l}}|\rightarrow {M}$. Such properties follow by exploiting the action of the (central extension of the) loop group $Map(S_{\theta (i)}^{(+)},SU(2))$ generated,
on the infinitesimal level, by the conserved currents 
\begin{gather}
J(\zeta (i))\doteq -\kappa \partial _{(i)}\widehat{h}_{i}\widehat{h}_{i}^{-1}
\\
\overline{J}(\overline{\zeta }(i))\doteq \kappa \widehat{h}_{i}^{-1}
\overline{\partial }_{(i)}\widehat{h}_{i},  \notag
\end{gather}
where $\partial _{(i)}\doteq \partial _{\zeta (i)}$. By writing $J(\zeta
(i))=J^{a}(\zeta (i))\mathbf{\sigma }_{a}$, we can introduce the
corresponding modes $J_{n}^{a}(i)$, from the Laurent expansion in each disk $
\delta _{\theta (i)}$, 
\begin{equation}
J^{a}(\zeta (i))=\sum_{n\in \mathbb{Z}}\zeta (i)^{-n-1}J_{n}^{a}(i),
\end{equation}
(and similarly for the modes $\overline{J}_{n}^{a}(i)$). The operator
product expansion of the currents $J^{a}(\zeta (i))J^{a}(\zeta ^{\prime }(i))
$, (with $\zeta (i)$ and $\zeta ^{\prime }(i)$ both in $\delta _{\theta (i)}$
) yields \cite{gawedzki} the commutation relations of an affine $\widehat{\mathfrak{su}}(2)$
algebra at the level $\kappa $, \emph{i.e.} 
\begin{equation}
\left[ J_{n}^{a}(i),J_{m}^{b}(i)\right] =\sqrt{-1}\varepsilon
_{abc}J_{n+m}^{c}(i)+\kappa n\delta _{ab}\delta _{n+m,0}.
\end{equation}
According to a standard procedure, we can then construct the Hilbert space $
\mathcal{H}_{(i)}$ associated with the  WZW fields $\widehat{h}_{i}$ by
considering unitary irreducible highest weight representations of the two
commuting copies of the current algebra $\widehat{\mathfrak{su}}(2)$ generated
by $J^{a}(\zeta (i))|_{S_{\theta (i)}^{(+)}}$ and $\overline{J^{a}}(
\overline{\zeta }(i))|_{S_{\theta (i)}^{(+)}}$. Such representations are
labelled by the level $\kappa $ and by the irreducible representations  of $
SU(2)$ with spin $0\leq \lambda (i)\leq \frac{\kappa }{2}$.
Note in particular that for $\kappa =1$ every highest weight representation of \ $\widehat{\mathfrak{su}}
(2)_{\kappa =1}$ also provides a representation of Virasoro algebra $Vir$
with central charge $c=1$. In such a case the representations of $\widehat{
\mathfrak{su}}(2)_{\kappa =1}$ can be decomposed into $\mathfrak{su}(2)\oplus Vir$,
and, up to Hilbert space completion, we can write 
\begin{equation}
\mathcal{H}_{(i)}=\bigoplus_{0\leq \lambda (i)\leq \frac{1}{2},0\leq n\leq
\infty }\left( V_{\mathfrak{su}(2)}^{(n+\lambda (i))}\otimes \overline{V}_{\mathfrak{
su}(2)}^{(n+\lambda (i))}\right) \otimes \left( \mathcal{H}_{(n+\lambda
(i))^{2}}^{Vir}\otimes \overline{\mathcal{H}}_{(n+\lambda
(i))^{2}}^{Vir}\right)   \label{Hilb1}
\end{equation}
where $V_{\mathfrak{su}(2)}^{(n+\lambda (i))}$ denotes the $(2\lambda (i)+1)$
-dimensional spin $\lambda (i)$ representation of $\mathfrak{su}(2)$, and $
\mathcal{H}_{(n+\lambda (i))^{2}}^{Vir}$ is the (irreducible highest weight)
representation of the Virasoro algebra of weight $(n+\lambda (i))^{2}$.
Since $0\leq \lambda (i)\leq \frac{1}{2}$, it is convenient to set 
\begin{equation}
j_{i}\doteq n+\lambda (i)\in \frac{1}{2}\mathbb{Z}^{+}
\end{equation}
(with $0\in \mathbb{Z}^{+}$), and rewrite (\ref{Hilb1}) as 
\begin{equation}
\mathcal{H}_{(i)}=\bigoplus_{j_{i},\widehat{j}_{i}\in \frac{1}{2}\mathbb{Z}
^{+}}\left( V_{\mathfrak{su}(2)}^{j_{i}}\otimes \overline{V}_{\mathfrak{su}(2)}^{
\widehat{j}_{i}}\right) \otimes \left( \mathcal{H}_{j_{i}^{2}}^{Vir}\otimes 
\overline{\mathcal{H}}_{\widehat{j}_{i}^{2}}^{Vir}\right) ,
\end{equation}
with $j_{i}+\widehat{j}_{i}$\ $\in \mathbb{Z}^{+}$, \cite{gaberdiel2}.\ \ Owing to
this particularly simple structure of the representation spaces $\mathcal{H}_{(i)}$, 
we shall limit our analysis to the case $\kappa =1$.

 Since
the boundary of $\partial M$ of the surface $M$ is defined by the disjoint
union $\bigsqcup S_{\theta (i)}^{(+)}$ and the boundary $\partial \Gamma $ \
of the ribbon graph $\Gamma $ \ is provided by $\bigsqcup S_{\theta
(i)}^{(-)}$, it follows that\ we can associate to both $\partial M$ and $
\partial \Gamma $ the Hilbert space 
\begin{equation}
\mathcal{H}(\partial M)\simeq \mathcal{H}(\partial \Gamma
)=\bigotimes_{i=1}^{N_{0}}\mathcal{H}_{(i)}.
\end{equation}
Let us denote by $\left| \widehat{h}(S_{\theta (i)}^{(+)})\right\rangle \in 
\mathcal{H}_{(i)}$ the Hilbert space state vector associated with the
boundary condition $\widehat{h}(S_{\theta (i)}^{(+)})$ on the $i$-th\
boundary component $S_{\theta (i)}^{(+)}$ of $M_{\partial }$. According to
the analysis of the previous section, the ribbon graph double $\widetilde{V}
_{M}$ generates a Schottky \ $M^{D}$\ double of the surface with cylindrical
boundaries $M_{\partial }$, ($M^{D}$ is the closed surface obtained by
identifying $M_{\partial }$ with another copy $M_{\partial }^{\prime }$ of $
M_{\partial }$ with opposite orientation along their common boundary $
\bigsqcup S_{\theta (i)}^{(+)}$). Such $M^{D}$\ carries an orientation
reversing involution 
\begin{equation}
\Upsilon :M^{D}\rightarrow M^{D},\;\Upsilon ^{2}=id
\end{equation}
that interchanges $M_{\partial }$ and $M_{\partial }^{\prime }$ and which
has the boundary $\bigsqcup S_{\theta (i)}^{(+)}$ as its fixed point set.
The request of preservation of conformal symmetry along $\bigsqcup S_{\theta
(i)}^{(+)}$ under the anticonformal involution $\Upsilon $ requires that the
state $\left| \widehat{h}(S_{\theta (i)}^{(+)})\right\rangle $ must satisfy
the glueing condition $(\mathbb{L}_{n}-\overline{\mathbb{L}}_{-n})\left| 
\widehat{h}(S_{\theta (i)}^{(+)})\right\rangle =0$, where, for $n\neq 0$, 
\begin{equation}
\mathbb{L}_{n}=\frac{1}{2+\kappa }\sum_{m=-\infty }^{\infty
}J_{n-m}^{a}J_{m}^{a},
\end{equation}
and similarly for $\overline{\mathbb{L}}_{-n}$. \ The glueing conditions
above can be solved mode by mode, and to each irreducible representation of
the Virasoro algebra $\mathcal{H}_{j_{i}^{2}}^{Vir}$ and its conjugate $
\overline{\mathcal{H}}_{\widehat{j}_{i}^{2}=j_{i}^{2}}^{Vir}$, labelled by
the given $j_{i}\doteq n+\lambda (i)\in \frac{1}{2}\mathbb{Z}^{+}$,\ we can
associate a set of conformal Ishibashi states parametrized by the $\mathfrak{su}
(2)$ representations $V_{\mathfrak{su}(2)}^{j_{i}}\otimes V_{\mathfrak{su}
(2)}^{j_{i}}$. Such states are usually denoted by 
\begin{equation}
\left. \left| j_{i};m,n\right\rangle \right\rangle ,\;\;m,n\in
(-j_{i},-j_{i}+1,...,j_{i}-1,j_{i}),
\end{equation}
and one can write \cite{gaberdiel}    
\begin{equation}
\left| \widehat{h}(S_{\theta (i)}^{(+)})\right\rangle =\frac{1}{2^{\frac{1}{4
}}}\sum_{j_{i};m,n}D_{m,n}^{j_{i}}(\widehat{h}(S_{\theta (i)}^{(+)}))\left.
\left| j_{i};m,n\right\rangle \right\rangle ,
\end{equation}
where 
\begin{align}
D_{m,n}^{j_{i}}(\widehat{h}(S_{\theta (i)}^{(+)}))& =\sum_{l=\max
(0,n-m)}^{\min (j_{i}-m,j_{i}+n)}\frac{\left[
(j_{i}+m)!(j_{i}-m)!(j_{i}+n)!(j_{i}-n)!\right] ^{\frac{1}{2}}}{
(j_{i}-m-l)!(j_{i}+n-l)!l!(m-n+l)!}\times  \\
& \times a^{j_{i}+n-l}\;d^{j_{i}-m-l}\;b^{l}\;c^{m-n+l},  \notag
\end{align}
is the $V_{\mathfrak{su}(2)}^{j_{i}}$-representation matrix associated with the $SU(2)$ element
\begin{equation}
\widehat{h}(S_{\theta (i)}^{(+)})=\left( 
\begin{array}{cc}
a & b \\ 
c & d
\end{array}
\right) \in C_{i}^{(+)},
\end{equation}
in the $C_{i}^{(+)}$ conjugacy class.

\subsection{The Quantum Amplitudes for the cylindrical ends}

\ \  With the above preliminary remarks along the way, let us consider explicitly the structure of the quantum amplitude associated with the  WZW model defined by the
action $S_{|P_{T_{l}}|}^{WZW}(\widehat{h})$. Formally, such an amplitude is provided by the
functional integral 
\begin{equation}
\left| \partial{M} ,\otimes_{i}\widehat{h}(S_{\theta (i)}^{(+)})\right\rangle =\int_{\{\widehat{
h}|_{S_{\theta (i)}^{(\pm )}}\in C_{i}^{(\pm
)}\}}e^{-S_{|P_{T_{l}}|}^{WZW}(\widehat{h})}D\widehat{h},
\label{amplit}
\end{equation}
where the integration is over maps $\widehat{h}$ satisfying the boundary conditions $\{\widehat{
h}|_{S_{\theta (i)}^{(\pm )}}\in C_{i}^{(\pm )}\}$, and where $D\widehat{
h}$ is the local product $\prod_{\zeta \in ((M;N_{0}),\mathcal{C})}d
\widehat{h}(\zeta )$ over $((M;N_{0}),\mathcal{C})$ of the $SU(2)$ Haar
measure. As the notation suggests,  the formal expression (\ref{amplit})
takes value in the Hilbert space $\mathcal{\ H}$. Let us recall that the fields $\widehat{
h}$ are constrained over \ the disjoint boundary components of $\partial
\Gamma $ \ to belong to the conjugacy classes $\{\widehat{h}|_{S_{\theta
(i)}^{(-)}}\in C_{i}^{(-)}\}$. This latter remark implies that the maps $\widehat{h}$ \ \ fluctuate
on the $N_{0}$ finite cylinders $\{\Delta
_{\theta (i)}^{\ast }\}$ wheras on the ribbon graph $\Gamma$ they are represented by boundary operators which mediate the changes in the boundary conditions on adjacent boundary components $\{\partial{\Gamma}_{i}\}$ of $\Gamma$. In order to exploit such a  factorization property of (\ref{amplit})
the first step is the computation of the amplitude, (for each given index $
i$),  for the cylinder $\Delta _{\theta (i)}^{\ast }$ with in and
out boundary conditions $\widehat{h}|_{S_{\theta (i)}^{(\pm )}}\in
C_{i}^{(\pm )}$, 
\begin{equation}
Z_{\Delta _{\theta (i)}^{\ast }}\doteq \int_{\widehat{h}|_{S_{i}^{(\pm )}}\in
C_{i}^{(\pm )}}e^{-S^{WZW}(\widehat{h};\Delta _{\theta (i)}^{\ast })}D
\widehat{h}
\end{equation}
where $S^{WZW}(\widehat{h};\Delta _{\theta (i)}^{\ast })$
is the restriction to $\Delta _{\theta (i)}^{\ast }$ of $
S_{|P_{T_{l}}|}^{WZW}(\widehat{h})$. If we introduce the Virasoro
operator $\mathbb{L}_{0}(i)$ defined by 
\begin{equation}
\mathbb{L}_{0}(i)=\frac{2}{2+\kappa }\sum_{m=0}^{\infty
}J_{-m}^{a}(i)J_{m}^{a}(i).
\end{equation}
and notice that $\mathbb{L}_{0}(i)+\overline{\mathbb{L}}_{0}(i)-\frac{c}{12}$
,  defines the Hamiltonian of the WZW theory on the cylinder $\Delta
_{\theta (i)}^{\ast }$, ($c=\frac{3\kappa }{2+\kappa }$ being the central
charge of the SU(2) WZW\ theory), then we can explicitly write
\begin{equation}
Z_{\Delta _{\theta (i)}^{\ast }}(\{C_{i}^{(\pm )}\})=\langle \widehat{h}(S_{\theta
(i)}^{(-)})|e^{-\frac{2\pi }{\theta (i)}(\mathbb{L}_{0}(i)+\overline{\mathbb{
L}}_{0}(i)-\frac{c}{12})}|\widehat{h}(S_{\theta (i)}^{(+)})\rangle ,
\label{annz}
\end{equation}
where 
$\langle \widehat{h}(S_{\theta (i)}^{(-)})|$ and $|\widehat{h}(S_{\theta
(i)}^{(+)})\rangle $ respectively denote the Hilbert space vectors associated with the boundary
conditions $h(S_{\theta (i)}^{(-)})$ and $h(S_{\theta (i)}^{(+)})$ and normalized to $\langle \widehat{h}(S_{\theta
(i)}^{(-)})||\widehat{h}(S_{\theta (i)}^{(+)})\rangle =1$
, (a normalization that follows from the fact that $\widehat{h}(S_{\theta (i)}^{(-)})$ and $
\widehat{h}(S_{\theta (i)}^{(+)})$ belong, by hypotheses, to the
conjugated 2-spheres $C_{i}^{(-)}$ and  $C_{i}^{(+)}$ in $SU(2)$).

\begin{figure}[ht]
\begin{center}
\includegraphics[scale=.4]{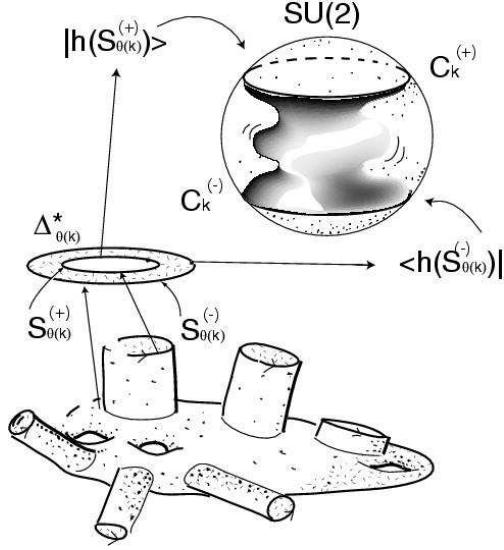}
\caption{A pictorial rendering of the set up for computing the quantum amplitudes for the cylindrical ends associated with the surface $\partial M$.}
\end{center}
\end{figure}

The computation of the annulus partition function (\ref{annz}) has been explicitly carried out \cite{gaberdiel} for the boundary $SU(2)$ CFT at level $\kappa=1$. We restrict our analysis to this particular case and  
if we apply the results of \cite{gaberdiel}, (see in particular eqn. (4.1) and the accompanying analysis) we get
\begin{multline}
Z_{\Delta _{\theta (i)}^{\ast }}(\{C_{i}^{\pm }\})= \\
=\frac{1}{\sqrt{2}}\sum_{j_i\in \frac{1}{2}\mathbb{Z}_{+}}\sum_{m,n}
(-1)^{m-n}D_{-m,-n}^{j_i}(\widehat{h}^{-1}(S_{\theta (i)}^{(-)}))D_{m,n}^{j_i}(\widehat{h}(S_{\theta
(i)}^{(+)}))\chi _{j_i^{2}}(e^{-\frac{4\pi }{\theta (i)}}),  
\end{multline}
where 
\begin{equation}
\chi _{j_i^{2}}(e^{-\frac{4\pi }{\theta (i)}})=\frac{e^{-\frac{4\pi }{\theta
(i)}j_i^{2}}-e^{-\frac{4\pi }{\theta (i)}(j_i+1)^{2}}}{\eta (e^{-\frac{4\pi }{
\theta (i)}})}.
\end{equation}
is the character of
the Virasoro highest weight representation, and 
\begin{equation}
\eta (q)\doteq q^{\frac{1}{24}}\prod_{n=1}^{\infty }(1-q^{n}),
\end{equation}
is the
Dedekind $\eta $-function. 

By diagonalizing we can consider $
h^{-1}(S_{\theta (i)}^{(-)})h(S_{\theta (i)}^{(+)})$ as an element of the maximal torus in $SU(2)$, i.e., we can write   
\begin{equation}
h^{-1}(S_{\theta (i)}^{(-)})h(S_{\theta (i)}^{(+)})=\left( 
\begin{array}{cc}
e^{4\pi \sqrt{-1}\lambda(i)} & 0 \\ 
0 & e^{-4\pi \sqrt{-1} \lambda(i)}
\end{array}
\right) ,
\end{equation}
and a representation-theoretic computation \cite{gaberdiel} eventually provides
\begin{equation}
Z_{\Delta _{\theta (i)}^{\ast }}(\{C_{i}^{\pm }\})=\frac{1}{\sqrt{2}}
\sum_{j\in \frac{1}{2}\mathbb{Z}_{+}}\cos (8\pi j_i\lambda(i))\frac{e^{-\frac{
4\pi }{\theta (i)}j_i^{2}}}{\eta (e^{-\frac{4\pi }{\theta (i)}})}.
\label{stringy}
\end{equation}
(Note that ${\alpha}$ in \cite{gaberdiel} corresponds to our $4\pi \sqrt{-1}\lambda(i)$, hence the presence of $\cos(8\pi j_i\lambda(i))$ in place of their $\cosh(2j_i\alpha(i))$). 

\bigskip

An important point to stress is that, according to the above analysis, the
partition function $Z_{\Delta _{\theta (i)}^{\ast }}(\{C_{i}^{\pm }\})$ can
be interpreted as the superposition over all possible $j_{i}$ channel
amplitudes 
\begin{equation}
\partial \Gamma _{i}\longmapsto A(j_{i})\doteq \frac{1}{\sqrt{2}}\cos
(8\pi j_{i}\lambda (i))\frac{e^{-\frac{4\pi }{\theta (i)}j_{i}^{2}}}{\eta (e^{-
\frac{4\pi }{\theta (i)}})}  \label{cellampl}
\end{equation}
that can be associated to the boundary component $\partial \Gamma _{i}$ of
the ribbon graph $\Gamma $. Such amplitudes can be interpreted as the
various $j_{i}=(n+\lambda (i))$, ($0\leq \lambda (i)\leq \frac{1}{2}$),
Virasoro (closed string) modes propagating along the cylinder $\Delta
_{\theta (i)}^{\ast }$.

\subsection{\protect\bigskip The Ribbon graph insertion operators}

In order to complete the picture, we need to discuss how the $N_{0}$
amplitudes $\{A(j_{i})\}$ defined by (\ref{cellampl}) interact along $\Gamma 
$. Such an interaction is described by boundary operators which mediate the
change in boundary conditions $|\widehat{h}(S_{\theta (p)}^{(+)})\rangle
_{\partial \Gamma _{p}}$ and $|\widehat{h}(S_{\theta (q)}^{(+)})\rangle
_{\partial \Gamma _{q}}$ between any two adjacent boundary components $
\partial \Gamma _{p}$ and $\partial \Gamma _{q}$, (note that the adjacent
boundaries of the ribbon graph are associated with adjacent cells $\rho
^{2}(p)$, $\rho ^{2}(q)$ of \ $|P_{T_{l}}|\rightarrow M$, and thus to the
edges $\sigma ^{1}(p,q)$ of the triangulation $|T_{l}|\rightarrow M$). In
particular, the coefficients of the operator product expansion (OPE),
describing the short-distance behavior of the boundary operators on adjacent 
$\partial \Gamma _{p}$ and $\partial \Gamma _{q}$, will keep tract of the
combinatorics associated with $|P_{T_{l}}|\rightarrow {M}$.

To this end, let us consider generic pairwise adjacent 2-cells $\rho ^{2}(p)$
, $\rho ^{2}(q)$ and $\rho ^{2}(r)$ in $|P_{T_{l}}|\rightarrow M$, and the
associated cyclically ordered 3-valent vertex $\rho ^{0}(p,q,r)\in
|P_{T_{l}}|\rightarrow M$. Let $\{U_{\rho ^{0}(p,q,r)},w\}$ the coordinate
neighborhood of such a vertex, and $\{U_{\rho ^{1}(p,q)},z\}$, $\{U_{\rho
^{1}(q,r)},z\}$, and $\{U_{\rho ^{1}(r,p)},z\}$ the neighborhoods of the
corresponding oriented edges, (the $z$'s appearing in distinct $\{U_{\rho
^{1}(\circ ,\bullet )},z\}$ are distinct). Consider the edge $\rho ^{1}(p,q)$
and two (infinitesimally neighboring) points $z_{1}=x_{1}+\sqrt{-1}y_{1}$
and $z_{2}=x_{2}+\sqrt{-1}y_{2}$, $\func{Re}z_{1}=\func{Re}z_{2}$, in the
corresponding $U_{\rho ^{1}(p,q)}$, with $x_{1}=x_{2}$. Thus, for $
y_{1}\rightarrow 0^{+}$ we approach $\partial \Gamma _{p}\cap \rho ^{1}(p,q)$
, whereas for $y_{2}\rightarrow 0^{-}$ we approach a point $\ \in \partial
\Gamma _{q}\cap \rho ^{1}(q,p)$.


Associated with the edge $\rho ^{1}(p,q)\ $
we have the two adjacent boundary conditions $|\widehat{h}(S_{\theta
(p)}^{(+)})\rangle _{\partial \Gamma _{p}}$, and $|\widehat{h}(S_{\theta
(q)}^{(+)})\rangle _{\partial \Gamma _{q}}$, respectively describing the
given values of the field $\widehat{h}$ on the two boundary components $
\partial \Gamma _{p}\cap \rho ^{1}(p,q)$ and $\partial \Gamma _{q}\cap \rho
^{1}(q,p)$ of $\rho ^{1}(p,q)$. At the points $z_{1},z_{2}\in $ $U_{\rho
^{1}(p,q)}$ we can consider the insertion of boundary operators $\psi
_{j_{(p,q)}}^{j_{q}j_{p}}(z_{1})$ and $\psi _{j_{(q,p)}}^{j_{p}j_{q}}(z_{2})$
mediating between the corresponding boundary conditions, \emph{i.e.} 
\begin{eqnarray}
&&\psi _{j_{(p,q)}}^{j_{q}j_{p}}(z_{1})|\widehat{h}(S_{\theta
(p)}^{(+)})\rangle _{\partial \Gamma _{p}}\underset{y_{1}\rightarrow 0^{+}}{=
}|\widehat{h}(S_{\theta (q)}^{(+)})\rangle _{\partial \Gamma _{q}},  \notag
\\
&& \\
&&\psi _{j_{(q,p)}}^{j_{p}j_{q}}(z_{2})|\widehat{h}(S_{\theta
(q)}^{(+)})\rangle _{\partial \Gamma _{q}}\underset{y_{2}\rightarrow 0^{-}}{=
}|\widehat{h}(S_{\theta (p)}^{(+)})\rangle _{\partial \Gamma _{p}}.  \notag
\end{eqnarray}
Note that $\psi _{j_{(p,q)}}^{j_{q}j_{p}}$ carries the single primary isospin label $
j_{(p,q)}$\ (also indicating the oriented edge $\rho ^{1}(p,q)$ where we are
inserting the operator), and the two additional isospin labels $j_{p}$ and $j_{q}$
indicating the two boundary conditions at the two portions of $\partial
\Gamma _{p}$ and $\partial \Gamma _{q}$\ adjacent to the insertion edge $
\rho ^{1}(p,q)$. \ Likewise, by considering the oriented edges $\rho
^{1}(q,r)$ and $\rho ^{1}(r,p)$, \ we can introduce the operators $\psi
_{j_{(r,q)}}^{j_{q}j_{r}}$, $\psi _{j_{(q,r)}}^{j_{r}j_{q}}$, $\psi
_{j_{(p,r)}}^{j_{r}j_{p}}$, and $\psi _{j_{(r,p)}}^{j_{p}j_{r}}$. In full
generality, we can rewrite the above definition explicitly in terms of the
adjacency matrix $B(\Gamma )$ of the ribbon graph $\Gamma $, 
\begin{equation}
B_{st}(\Gamma )=\left\{ 
\begin{tabular}{lll}
$1$ & if & $\rho ^{1}(s,t)$ is an edge of $\Gamma $ \\ 
&  &  \\ 
$0$ &  & otherwise
\end{tabular}
\right. ,
\end{equation}
according to 
\begin{equation}
\psi _{j_{(p,q)}}^{j_{q}j_{p}}(z_{1})|\widehat{h}(S_{\theta
(p)}^{(+)})\rangle _{\partial \Gamma _{p}}\underset{y_{1}\rightarrow 0^{+}}{=
}B_{pq}(\Gamma )|\widehat{h}(S_{\theta (q)}^{(+)})\rangle _{\partial \Gamma
_{q}}.
\end{equation}
Any\ such boundary operator, say $\psi _{j_{(p,q)}}^{j_{q}j_{p}}$, is a
primary field (under the action of Virasoro algebra) of conformal dimension $
H_{j_{(p,q)}}$, and they are all characterized \cite{sagnotti}, \cite{lewellen}, \cite{felder} by the following properties
dictated by conformal invariance (in the corresponding coordinate
neighborhood $U_{\rho ^{1}(p,q)}$) 
\begin{gather}
\langle 0|\psi _{j_{(p,q)}}^{j_{q}j_{p}}(z_{1})|0\rangle =0,\;\;\langle 
\widehat{h}(S_{\theta (p)}^{(-)})|\mathbb{I}^{j_{p}j_{p}}|\widehat{h}
(S_{\theta (p)}^{(+)})\rangle \;\!=a^{j_{p}j_{p}},  \notag \\
\label{twopoints} \\
\langle 0|\psi _{j_{(p,q)}}^{j_{q}j_{p}}(z_{1})\psi
_{j_{(q,p)}}^{j_{p}j_{q}}(z_{2})|0\rangle
=b_{j_{(p,q)}}^{j_{q}j_{p}}|z_{1}-z_{2}|^{-2H_{j_{(p,q)}}}\delta
_{j_{(p,q)}j_{(q,p)}},  \notag
\end{gather}
where $\mathbb{I}^{j_{p}j_{p}}$ is the identity operator, and where $
a^{j_{p}j_{p}}$ and $b_{j_{(p,q)}}^{j_{q}j_{p}}$ are normalization factors.
In particular, the parameters $b_{j_{(p,q)}}^{j_{q}j_{p}}$ define the
normalization of the two-points function. Note that \cite{sagnotti} for $SU(2)$ the $
b_{j_{(p,q)}}^{j_{q}j_{p}}$ are such that $
b_{j_{(p,q)}}^{j_{q}j_{p}}=b_{j_{(q,p)}}^{j_{p}j_{q}}(-1)^{2j_{(p,q)}}$, and
are (partially) constrained by the OPE of the $\psi _{j_{(p,q)}}^{j_{q}j_{p}}
$. As customary in boundary CFT, we leave such a normalization factors
dependence explicit in what follows.

\begin{figure}[ht]
\begin{center}
\includegraphics[scale=.5]{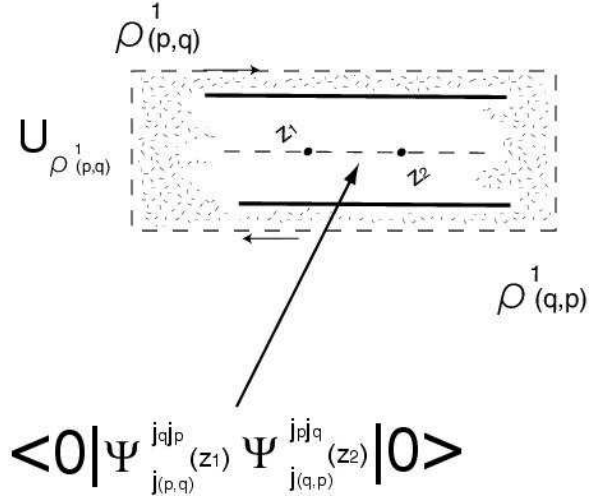}
\caption{The insertion of boundary operators $\psi _{j_{(p,q)}}^{j_{q}j_{p}}$
 in the complex coordinate neighborhood $U_{\rho ^{1}(p,q)}$, giving rise to the two-point function in the corresponding oriented edge $\rho^{1}(p,q)$.}
\end{center}
\end{figure}

In order to discuss the properties of \ the $\psi _{j_{(p,q)}}^{j_{q}j_{p}}$
, let us extend the (edges) coordinates $z$\ \ to the unit disk \ $U_{\rho
^{0}(p,q,r)}$ associated to the generic vertex $\rho ^{0}(p,q,r)$, and
denote by 
\begin{eqnarray}
w_{p} &=&\frac{\varepsilon }{3}e^{\frac{1}{2}\pi \sqrt{-1}}\in U_{\rho
^{0}(p,q,r)}\cap U_{\rho ^{1}(p,q)}  \notag \\
w_{q} &=&\frac{\varepsilon }{2}e^{\frac{7}{6}\pi \sqrt{-1}}\in U_{\rho
^{0}(p,q,r)}\cap U_{\rho ^{1}(q,r)}  \label{wcoord} \\
w_{r} &=&\varepsilon e^{\frac{11}{6}\pi \sqrt{-1}}\in U_{\rho
^{0}(p,q,r)}\cap U_{\rho ^{1}(r,p)}  \notag
\end{eqnarray}
the coordinates of three points in an $\varepsilon $- neighborhood ($
0<\varepsilon <1$) of the vertex $w=0$, (fractions of $\varepsilon $ are
introduced for defining a radial ordering; note also that by exploting the
coordinate changes (\ref{glue1}), one can easily map such points in the upper half
planes associated with the edge complex variables $z$ corresponding to $
U_{\rho ^{1}(p,q)}$, $U_{\rho ^{1}(q,r)}$, and $U_{\rho ^{1}(r,p)}$, and
formulate the theory in a more conventional fashion). To
these points we associate the insertion of boundary operators $\psi
_{j_{(r,p)}}^{j_{p}j_{r}}(w_{r})$, $\psi _{j_{(q,r)}}^{j_{r}j_{q}}(w_{q})$, $
\psi _{j_{(p,q)}}^{j_{q}j_{p}}(w_{p})$ which pairwise mediate among the
boundary conditions $|\widehat{h}(S_{\theta (p)}^{(+)})\rangle $, $|\widehat{
h}(S_{\theta (q)}^{(+)})\rangle $, and $|\widehat{h}(S_{\theta
(r)}^{(+)})\rangle $. The behavior of such insertions at the vertex $\rho
^{0}(p,q,r)$, (\emph{i.e.}, as $\varepsilon \rightarrow 0$), is described by
the following OPEs (see \cite{lewellen}, \cite{sagnotti})
\begin{gather}
\psi _{j_{(r,p)}}^{j_{p}j_{r}}(w_{r})\psi _{j_{(q,r)}}^{j_{r}j_{q}}(w_{q})= 
\notag \\
\label{OPE} \\
=\sum_{j}C_{j_{(r,p)}j_{(q,r)}j}^{j_{p}j_{r}j_{q}}|w_{r}-w_{q}|^{H_{j}-H_{j_{(r,p)}}-H_{j_{(q,r)}}}(\psi _{j}^{j_{p}j_{q}}(w_{q})+...),
\notag
\end{gather}
\begin{gather}
\psi _{j_{(q,r)}}^{j_{r}j_{q}}(w_{q})\psi _{j_{(p,q)}}^{j_{q}j_{p}}(w_{p})= 
\notag \\
\\
=\sum_{j}C_{j_{(q,r)}j_{(p,q)}j}^{j_{r}j_{q}j_{p}}|w_{q}-w_{p}|^{H_{j}-H_{j_{(q,r)}}-H_{j_{(p,q)}}}(\psi _{j}^{j_{r}j_{p}}(w_{p})+...),
\notag
\end{gather}
\begin{gather}
\psi _{j_{(p,q)}}^{j_{q}j_{p}}(w_{p})\psi _{j_{(r,p)}}^{j_{p}j_{r}}(w_{r})= 
\notag \\
\\
=\sum_{j}C_{j_{(p,q)}j_{(r,p)}j}^{j_{q}j_{p}j_{r}}|w_{p}-w_{r}|^{H_{j}-H_{j_{(p,q)}}-H_{j_{(r,p)}}}(\psi _{j}^{j_{q}j_{p}}(w_{r})+...),
\notag
\end{gather}
where the dots stand for higher order corrections in $|w_{\circ }-w_{\bullet
}|$, the $H_{J_{...}}$ are the conformal weights of the corresponding
boundary operators, and the $C_{j_{(r,p)}j_{(q,r)}j}^{j_{p}j_{r}j_{q}}$\ are
the OPE structure constants.

\begin{figure}[ht]
\begin{center}
\includegraphics[scale=.5]{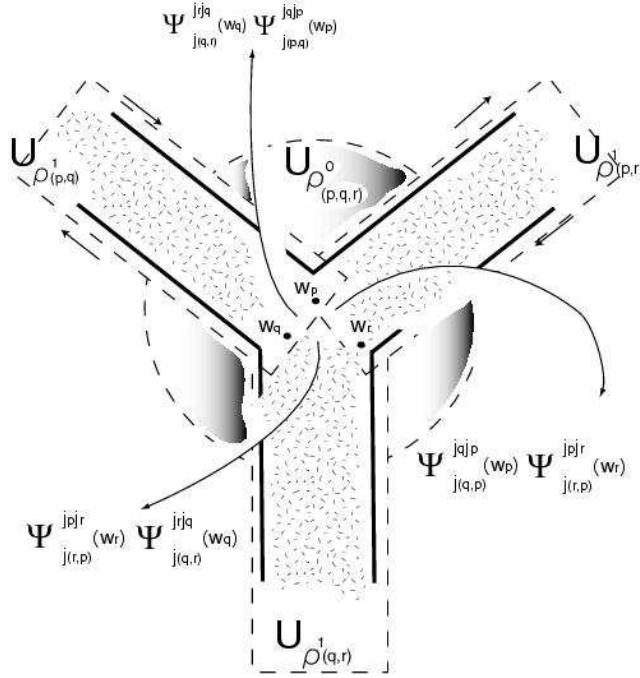}
\caption{The OPEs between the boundary operators around a given vertex $\rho^{0}(p,q,r)$ in the corresponding complex coordinates neighborhoods $U_{\rho ^{0}(p,q,r)}$, $U_{\rho ^{1}(p,q)}$, etc..}
\end{center}
\end{figure}

As is well known \cite{sagnotti}, the parameters $
b_{j_{(p,q)}}^{j_{q}j_{p}}$ and the constants $
C_{j_{(r,p)}j_{(q,r)}j}^{j_{p}j_{r}j_{q}}$ are not independent. In our
setting this is a consequence of the fact that to the oriented vertex $\rho
^{0}(p,q,r)$ we can associate a three-point function which must be invariant
under cyclic permutations, \emph{i.e.} 
\begin{gather}
\langle \psi _{j_{(r,p)}}^{j_{p}j_{r}}(w_{r})\psi
_{j_{(q,r)}}^{j_{r}j_{q}}(w_{q})\psi _{j_{(p,q)}}^{j_{q}j_{p}}(w_{p})\rangle
=\langle \psi _{j_{(p,q)}}^{j_{q}j_{p}}(w_{r})\psi
_{j_{(r,p)}}^{j_{p}j_{r}}(w_{q})\psi _{j_{(q,r)}}^{j_{r}j_{q}}(w_{p})\rangle
=  \notag \\
\\
=\langle \psi _{j_{(q,r)}}^{j_{r}j_{q}}(w_{r})\psi
_{j_{(p,q)}}^{j_{q}j_{p}}(w_{q})\psi _{j_{(r,p)}}^{j_{p}j_{r}}(w_{p})\rangle
.  \notag
\end{gather}
By using the boundary OPE (\ref{OPE}), each term can be computed in two
distinct ways, \emph{e.g.}, by denoting with $\underbrace{}$\ an OPE
pairing, we must have 
\begin{equation}
\langle \underbrace{\psi _{j_{(r,p)}}^{j_{p}j_{r}}(w_{r})\psi
_{j_{(q,r)}}^{j_{r}j_{q}}(w_{q})}\psi
_{j_{(p,q)}}^{j_{q}j_{p}}(w_{p})\rangle =\langle \psi
_{j_{(r,p)}}^{j_{p}j_{r}}(w_{r})\underbrace{\psi
_{j_{(q,r)}}^{j_{r}j_{q}}(w_{q})\psi _{j_{(p,q)}}^{j_{q}j_{p}}(w_{p})}
\rangle 
\end{equation}
which (by exploiting (\ref{twopoints})) in the limit $w\rightarrow 0$\ \
provides 
\begin{equation}
C_{j_{(r,p)}j_{(q,r)}j_{(p,q)}}^{j_{p}j_{r}j_{q}}b_{j_{(q,p)}}^{j_{p}j_{q}}=C_{j_{(q,r)}j_{(p,q)}j_{(r,p)}}^{j_{r}j_{q}j_{p}}b_{j_{(r,p)}}^{j_{p}j_{r}},
\label{Cval}
\end{equation}
(note that the Kronecker $\delta $ in (\ref{twopoints}) implies that 
$j_{(q,p)}=j_{(p,q)}$, etc. ). From the OPE evaluation of the remaining
two three-points function one similarly obtains 
\begin{eqnarray}
C_{j_{(p,q)}j_{(r,p)}j_{(q,r)}}^{j_{q}j_{p}j_{r}}b_{j_{(r,q)}}^{j_{q}j_{r}}
&=&C_{j_{(r,p)}j_{(q,r)}j_{(p,q)}}^{j_{p}j_{r}j_{q}}b_{j_{(p,q)}}^{j_{q}j_{p}},
\notag \\
&& \\
C_{j_{(q,r)}j_{(p,q)}j_{(r,p)}}^{j_{r}j_{q}j_{p}}b_{j_{(p,r)}}^{j_{r}j_{p}}
&=&C_{j_{(p,q)}j_{(r,p)}j_{(q,r)}}^{j_{q}j_{p}j_{r}}b_{j_{(q,r)}}^{j_{r}j_{q}}.
\notag
\end{eqnarray}
Since 
\begin{eqnarray}
b_{j_{(q,p)}}^{j_{p}j_{q}} &=&b_{j_{(p,q)}}^{j_{q}j_{p}}(-1)^{2j_{(p,q)}}, 
\notag \\
b_{j_{(r,p)}}^{j_{p}j_{r}} &=&b_{j_{(p,r)}}^{j_{r}j_{p}}(-1)^{2j_{(p,r)}}, \\
b_{j_{(q,r)}}^{j_{r}j_{q}} &=&b_{j_{(r,q)}}^{j_{q}j_{r}}(-1)^{2j_{(r,q)}}, 
\notag
\end{eqnarray}
one eventually gets 
\begin{eqnarray}
C_{j_{(r,p)}j_{(q,r)}j_{(p,q)}}^{j_{p}j_{r}j_{q}}b_{j_{(q,p)}}^{j_{p}j_{q}}
&=&(-1)^{2j_{(q,p)}}C_{j_{(p,q)}j_{(r,p)}j_{(q,r)}}^{j_{q}j_{p}j_{r}}b_{j_{(q,r)}}^{j_{r}j_{q}},
\notag \\
&&  \notag \\
C_{j_{(p,q)}j_{(r,p)}j_{(q,r)}}^{j_{q}j_{p}j_{r}}b_{j_{(r,q)}}^{j_{q}j_{r}}
&=&(-1)^{2j_{(r,q)}}C_{j_{(q,r)}j_{(p,q)}j_{(r,p)}}^{j_{r}j_{q}j_{p}}b_{j_{(r,p)}}^{j_{p}j_{r}},
\\
&&  \notag \\
C_{j_{(q,r)}j_{(p,q)}j_{(r,p)}}^{j_{r}j_{q}j_{p}}b_{j_{(p,r)}}^{j_{r}j_{p}}
&=&(-1)^{2j_{(p,r)}}C_{j_{(r,p)}j_{(q,r)}j_{(p,q)}}^{j_{p}j_{r}j_{q}}b_{j_{(p,q)}}^{j_{q}j_{p}},
\notag
\end{eqnarray}
which are the standard relation between the OPE parameters and the normalization of the 2-points function for boundary $SU(2)$ insertion operators, \cite{sagnotti}. Such a lengthy (and slightly pedantic) analysis is necessary to show that our association of boundary insertion operators $\psi
_{j_{(r,p)}}^{j_{p}j_{r}}$,  to the edges of the ribbon graph $\Gamma $ is actually consistent with $SU(2)$ boundary CFT, in particular that geometrically the correlator $\langle \psi
_{j_{(r,p)}}^{j_{p}j_{r}}(w_{r})\psi _{j_{(q,r)}}^{j_{r}j_{q}}(w_{q})\psi
_{j_{(p,q)}}^{j_{q}j_{p}}(w_{p})\rangle $ \ \ is associated with the three
mutually adjacent boundary components $\partial \Gamma _{p}$, $\partial
\Gamma _{q}$, and $\partial \Gamma _{r}$ of the ribbon graph $\Gamma $. More
generally, let us consider four mutually adjacent boundary components $
\partial \Gamma _{p}$, $\partial \Gamma _{q}$, $\partial \Gamma _{r}$, and $
\partial \Gamma _{s}$. Their adjacency relations can be organized in two
distinct ways labelled by the distinct two vertices they generate: if $
\partial \Gamma _{p}$ is adjacent to $\partial \Gamma _{r}$ then \ we have
the two vertices $\rho ^{0}(p,q,r)$ and $\rho ^{0}(p,r,s)$ connected by the
edge $\rho ^{1}(p,r)$; conversely, if $\partial \Gamma _{q}$ is adjacent to $
\partial \Gamma _{s}$ then \ we have the two vertices $\rho ^{0}(p,q,s)$ and 
$\rho ^{0}(q,r,s)$ connected by the edge $\rho ^{1}(q,s)$. \ It follows that
the correlation function of the corresponding four boundary operators, $
\langle \psi _{j_{(s,p)}}^{j_{p}j_{s}}\psi _{j_{(r,s)}}^{j_{s}j_{r}}\psi
_{j_{(q,r)}}^{j_{r}j_{q}}\psi _{j_{(p,q)}}^{j_{q}j_{p}}\rangle $, can be
evaluated by exploiting the ($(S)$-channel) factorization associated with
the coordinate neighborhood $\{U_{\rho ^{1}(r,p)},z^{(S)}\}$, or,
alternatively, by exploiting the ($(T)$-channel) factorization associated
with $\{U_{\rho ^{1}(q,s)},z^{(T)}\}$. 

\begin{figure}[ht]
\begin{center}
\includegraphics[scale=.6]{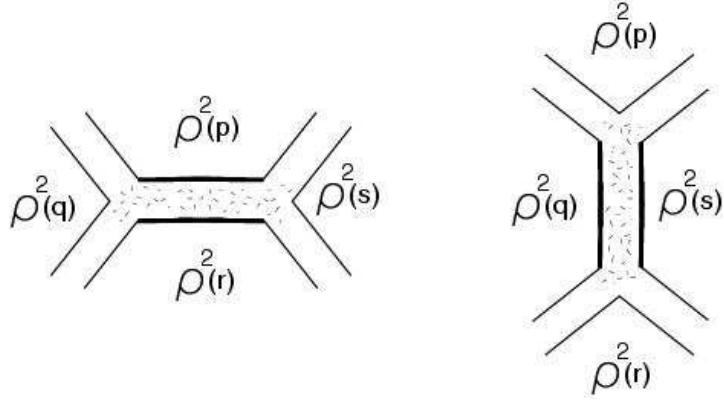}
\caption{The dual channels in evaluating the correlation function of the four boundary operators corresponding to the four boundary components involved.}
\end{center}
\end{figure}

From the observation that both such
expansions must yield the same result, it is possible \cite{felder} to directly relate the
OPE coefficients $C_{j_{(r,p)}j_{(q,r)}j_{(p,q)}}^{j_{p}j_{r}j_{q}}$ with
the fusion matrices $F_{j_{r}j_{(p,q)}}\left[\begin{array}{cc}
j_p & j_q\\
j_{(r,p)} & j_{(q,r)}\end{array}\right]$ which express the crossing duality between
four-points conformal blocks. Recall that for WZW models the fusion ring can
be identified with the character ring of the quantum deformation $\mathcal{U}
_{Q}(\mathbf{g})$ of \ the enveloping algebra of $\mathbf{g}$\ evaluated at
the root of unity given by $Q=e^{\pi \sqrt{-1}/(\kappa +h^{\vee })}$\ (where 
$h^{\vee }$ is the dual Coxeter number and $\kappa $ is the level of the
theory). In other words, for WZW models, the fusion\ \ matrices are the $6j$
-symbols of the corresponding (quantum) group. From such remarks, it follows
that in our case (\emph{i.e.}, for $\kappa =1$, $h^{\vee }=2$) the structure
constants $C_{j_{(r,p)}j_{(q,r)}j_{(p,q)}}^{j_{p}j_{r}j_{q}}$ are suitable
entries \cite{gaume} of the $6j$-symbols of the quantum group $SU(2)_{Q=e^{\frac{\pi }{3}
\sqrt{-1}}}$, $\emph{i.e.}$
\begin{equation}
C_{j_{(r,p)}j_{(q,r)}j_{(p,q)}}^{j_{p}j_{r}j_{q}}=\left\{ 
\begin{array}{ccc}
j_{(r,p)} & j_{p} & j_{r} \\ 
j_{q} & j_{(q,r)} & j_{(p,q)}
\end{array}
\right\} _{Q=e^{\frac{\pi }{3}\sqrt{-1}}}  \label{seigei}
\end{equation}

\subsection{The partition function.}

The final step in our construction is to uniformize the local coordinate
representation of the ribbon graph $\Gamma $ with the cylindrical metric $
\{|\phi (i)|\}$, defined by the quadratic differential $\{\phi (i)\}$. In
such a framework, there is a natural prescription for associating to the
resulting metric ribbon graph $(\Gamma ,\{|\phi (i)|\})$ a factorization of
the correlation functions of $\ $the $N_{1}$ insertion operators $\{\psi
_{j_{(p,q)}}^{j_{q}j_{p}}\}$, (recall that $N_{1}$ is the number of edges of 
$\Gamma $). Explicitly, for the generic vertex \ $\rho ^{0}(p,q,r)$, let \ $
z_{p}^{(0)}\in U_{\rho ^{1}(p,q)}\cap U_{\rho ^{0}(p,q,r)}$, \ $
z_{q}^{(0)}\in U_{\rho ^{1}(q,r)}\cap U_{\rho ^{0}(p,q,r)}$, and \ $
z_{r}^{(0)}\in U_{\rho ^{1}(r,p)}\cap U_{\rho ^{0}(p,q,r)}$ respectively
denote the coordinates of the points $w_{p}$, $w_{q}$, and $w_{r}$ (see (\ref
{wcoord})) in the respective edge uniformizations, and for notational
purposes, let us set, (in an $\varepsilon $-neighborhood of $z_{\rho
^{0}(p,q,r)}=0\in U_{\rho ^{0}(p,q,r)}$), 
\begin{eqnarray}
\psi _{j_{(r,p)}}^{j_{p}j_{r}}(\rho ^{0}(p,q,r)) &\doteq &\psi
_{j_{(r,p)}}^{j_{p}j_{r}}(z_{r}^{(0)}),  \notag \\
\psi _{j_{(q,r)}}^{j_{r}j_{q}}(\rho ^{0}(p,q,r)) &\doteq &\psi
_{j_{(q,r)}}^{j_{r}j_{q}}(z_{q}^{(0)}), \\
\psi _{j_{(p,q)}}^{j_{q}j_{p}}(\rho ^{0}(p,q,r)) &\doteq &\psi
_{j_{(p,q)}}^{j_{q}j_{p}}(z_{p}^{(0)}).  \notag
\end{eqnarray}
Let us consider, (in the limit $\varepsilon \rightarrow 0$ ),\ the
correlation function 
\begin{gather}
\left\langle \bigotimes_{i=1}^{N_{0}(T)}\partial \Gamma _{i};\otimes
j_{i}\right\rangle \doteq   \notag \\
\\
\doteq \left\langle \prod_{\{\rho ^{0}(p,q,r)\}}^{N_{2}(T)}\psi
_{j_{(r,p)}}^{j_{p}j_{r}}(\rho ^{0}(p,q,r))\psi
_{j_{(q,r)}}^{j_{r}j_{q}}(\rho ^{0}(p,q,r))\psi
_{j_{(p,q)}}^{j_{q}j_{p}}(\rho ^{0}(p,q,r))\right\rangle ,  \notag
\end{gather}
where \ the product runs over the $N_{2}(T)$ vertices $\{\rho ^{0}(p,q,r)\}$
\ of $\Gamma $.\ We can factorize it along the $N_{1}(T)$ channels generated
by the edge cordinate neighborhoods $\{U_{\rho ^{1}(p,q)}\}$ according to 
\begin{gather}
\left\langle \bigotimes_{i=1}^{N_{0}(T)}\partial \Gamma _{i};\otimes
j_{i}\right\rangle =  \notag \\
\\
=\sum_{\{j_{(r,p)}\}}\prod_{\{\rho ^{0}(p,q,r)\}}^{N_{2}(T)}\underset{\rho
^{0}(p,q,r)}{\left\langle \psi _{j_{(r,p)}}^{j_{p}j_{r}}\psi
_{j_{(q,r)}}^{j_{r}j_{q}}\psi _{j_{(p,q)}}^{j_{q}j_{p}}\right\rangle }
\prod_{\{\rho ^{1}(p,r)\}}^{N_{1}(T)}\underset{\rho ^{1}(p,r)}{\left\langle
\psi _{j_{(r,p)}}^{j_{p}j_{r}}\psi _{j_{(p,r)}}^{j_{r}j_{p}}\right\rangle } 
\notag
\end{gather}
where we have set 
\begin{gather}
\underset{\rho ^{0}(p,q,r)}{\left\langle \psi _{j_{(r,p)}}^{j_{p}j_{r}}\psi
_{j_{(q,r)}}^{j_{r}j_{q}}\psi _{j_{(p,q)}}^{j_{q}j_{p}}\right\rangle }\doteq 
\notag \\
\\
\doteq \left\langle \psi _{j_{(r,p)}}^{j_{p}j_{r}}(\rho ^{0}(p,q,r))\psi
_{j_{(q,r)}}^{j_{r}j_{q}}(\rho ^{0}(p,q,r))\psi
_{j_{(p,q)}}^{j_{q}j_{p}}(\rho ^{0}(p,q,r))\right\rangle ,  \notag
\end{gather}
\begin{equation}
\underset{\rho ^{1}(p,r)}{\left\langle \psi _{j_{(r,p)}}^{j_{p}j_{r}}\psi
_{j_{(p,r)}}^{j_{r}j_{p}}\right\rangle }\doteq \left\langle \psi
_{j_{(r,p)}}^{j_{p}j_{r}}(\rho ^{0}(p,q,r))\psi
_{j_{(p,r)}}^{j_{r}j_{p}}(\rho ^{0}(p,r,s))\right\rangle ,
\end{equation}
and where the summation runs over all $N_{1}(T)$ primary highest weight
representation $\widehat{\mathfrak{su}}(2)_{\kappa =1}$,\ labelling the
intermediate edge channels $\{j_{(r,p)}\}$. Note that according to (\ref
{twopoints}) we can write 
\begin{equation}
\underset{\rho ^{1}(p,r)}{\left\langle \psi _{j_{(r,p)}}^{j_{p}j_{r}}\psi
_{j_{(p,r)}}^{j_{r}j_{p}}\right\rangle }
=b_{j_{(r,p)}}^{j_{p}j_{r}}L(p,r)^{-2H_{j_{(r,p)}}},
\label{critedge}
\end{equation}
(recall that $j_{(r,p)}=j_{(p,r)}$), where $L(p,r)$ denotes the length of
the edge $\rho ^{1}(p,r)$ in the uniformization $(U_{\rho ^{1}(p,r)},\{|\phi
(i)|\})$. Moreover, \ since (see (\ref{Cval})) 
\begin{equation}
\underset{\rho ^{0}(p,q,r)}{\left\langle \psi _{j_{(r,p)}}^{j_{p}j_{r}}\psi
_{j_{(q,r)}}^{j_{r}j_{q}}\psi _{j_{(p,q)}}^{j_{q}j_{p}}\right\rangle }
=C_{j_{(r,p)}j_{(q,r)}j_{(p,q)}}^{j_{p}j_{r}j_{q}}b_{j_{(q,p)}}^{j_{p}j_{q}},
\end{equation}
we get for the boundary operator correlator associated with the 
ribbon graph $\Gamma $ the expression 
\begin{gather}
\left\langle \bigotimes_{i=1}^{N_{0}(T)}\partial \Gamma _{i};\otimes
j_{i}\right\rangle =  \notag \\
\\
=\sum_{\{j_{(r,p)}\}}\prod_{\{\rho
^{0}(p,q,r)
\}}^{N_{2}(T)}C_{j_{(r,p)}j_{(q,r)}j_{(p,q)}}^{j_{p}j_{r}j_{q}}b_{j_{(q,p)}}^{j_{p}j_{q}}\prod_{\{\rho ^{1}(p,r)\}}^{N_{1}(T)}b_{j_{(r,p)}}^{j_{p}j_{r}}L(p,r)^{-2H_{j_{(r,p)}}}.
\notag
\end{gather}


By identifying each $C_{j_{(r,p)}j_{(q,r)}j_{(p,q)}}^{j_{p}j_{r}j_{q}}$ with
the corresponding $6j$-symbol, and observing that each normalization factor $
b_{j_{(q,p)}}^{j_{p}j_{q}}$ occurs exactly twice, we eventually obtain 
\begin{gather}
\left\langle \bigotimes_{i=1}^{N_{0}(T)}\partial \Gamma _{i};\otimes
j_{i}\right\rangle =  \notag \\
\label{graphA} \\
\sum_{\{j_{(r,p)}\}}\prod_{\{\rho ^{0}(p,q,r)\}}^{N_{2}(T)}
\left\{ 
\begin{array}{ccc}
j_{(r,p)} & j_{p} & j_{r} \\ 
j_{q} & j_{(q,r)} & j_{(p,q)}
\end{array}
\right\} _{Q=e^{\frac{\pi }{3}\sqrt{-1}}}\prod_{\{\rho
^{1}(p,r)\}}^{N_{1}(T)}\left( b_{j_{(r,p)}}^{j_{p}j_{r}}\right)
^{2}L(p,r)^{-2H_{j_{(r,p)}}}.  \notag
\end{gather}
As the notation suggests, \ such a correlator has a residual
dependence on the representation labels $\{j_{i}\}$. In other words, it can
be considered as an element of the tensor product $\mathcal{H}(\partial
\Gamma )=$ $\otimes _{i=1}^{N_{0}(T)}\mathcal{H}_{(i)}$. It is then natural
to interpret its evaluation over the amplitudes $\{A(j_{i})\}$ defined by (
\ref{cellampl}) as the partition function $Z^{WZW}(|P_{T_{l}}|,\{\widehat{h}
(S_{\theta (i)}^{(+)})\})$ associated with the quantum amplitude (\ref{amplit}),
and describing the $SU(2)$ WZW model (at level $\kappa =1$) on a random
Regge polytope $|P_{T_{l}}|\rightarrow M$. By inserting the $N_{0}(T)$
amplitudes $\{A(j_{i})\}$ into (\ref{graphA}), and summing over all possible
representation indices $\{j_{p}\}$ we immediately get 
\begin{gather}
Z^{WZW}(|P_{T_{l}}|,\{\widehat{h}(S_{\theta (i)}^{(+)})\})=  \notag \\
\notag \\
=\left( \frac{1}{\sqrt{2}}\right) ^{N_{0}(T)}\sum_{\{j_{p}\in \frac{1}{2}
\mathbb{Z}_{+}\}}\sum_{\{j_{(r,p)}\}}\prod_{\{\rho
^{0}(p,q,r)\}}^{N_{2}(T)}\left\{ 
\begin{array}{ccc}
j_{(r,p)} & j_{p} & j_{r} \\ 
j_{q} & j_{(q,r)} & j_{(p,q)}
\end{array}
\right\} _{Q=e^{\frac{\pi }{3}\sqrt{-1}}}\cdot   \label{FinPart} \\
\notag \\
\cdot \prod_{\{\rho ^{1}(p,r)\}}^{N_{1}(T)}\left(
b_{j_{(r,p)}}^{j_{p}j_{r}}\right) ^{2}L(p,r)^{-2H_{j_{(r,p)}}}\cos
(8\pi j_{p}\lambda (i))\frac{e^{-\frac{4\pi }{\theta (i)}j_{p}^{2}}}{\eta (e^{-
\frac{4\pi }{\theta (i)}})},  \notag
\end{gather}
where the summation $\sum_{\{j_{p}\in \frac{1}{2}\mathbb{Z}_{+}\}}$ is over
all possible $N_{0}(T)$ channels $j_{p}$ describing the Virasoro (closed
string) modes propagating along the cylinders $\{\Delta _{\theta (p)}^{\ast
}\}_{p=1}^{N_{0}(T)}$.
This is the partition function of our WZW model on a random Regge triangulation. The WZW fields are still present through their boundary labels $\lambda(i)$, (which can take the values 
$0,1/2$), wheras the metric geometry of the polytope enters
explicitly both with the edge-length terms $
L(p,r)^{-2H_{j_{(r,p)}}}$ and with the conical angle factors $\frac{e^{-
\frac{4\pi }{\theta (i)}j_{p}^{2}}}{\eta (e^{-\frac{4\pi }{\theta (i)}})}$. The expression of 
$Z^{WZW}(|P_{T_{l}}|,\{\widehat{h}
(S_{\theta (i)}^{(+)})\})$, also shows the mechanism through which the $SU(2)$ fields couple with
simplicial curvature: the coupling amplitudes $\{A(j_{i})\}$ can be
interpreted as describing a closed string emitted by $\partial \Gamma
_{i}\simeq S_{\theta (i)}^{(-)}$, or rather by the $\overline{S_{\theta
(i)}^{2}}$\ \ brane image of this boundary component in $SU(2)$, and
absorbed by the brane $S_{\theta (i)}^{2}$ image of the outer boundary $
S_{\theta (i)}^{(+)}$, (the curvature carrying vertex). This exchange of
closed strings between $2$-branes in \ $SU(2)\simeq S^{3}$ describes the
interaction of the quantum $SU(2)$ field with the classical gravitational
background associated with the edge-length assignments $\{L(p,r)\}$, and
with the deficit angles $\{\varepsilon (i)\doteq 2\pi -\theta (i)\}$.

\section{\protect\bigskip Concluding remarks}

We note on passing that, in the above framework, 2D gravity can be promoted to a dynamical role by summing (\ref
{FinPart}) over all possible Regge polytopes (\emph{i.e.}, over all possible
metric ribbon graphs $\{\Gamma ,\{L(p,r)\}\}$). It is clear, from the
edge-lenght dependence in (\ref{FinPart}), that the formal Regge functional
measure $\propto \prod_{\{\rho ^{1}(p,r)\}}dL(p,r)$, involved in such a
summation, inherits an anomalous scaling related to the presence of the weighting
factor (to be summed over all isospin channels $j(r,p)$)

\begin{equation}
\prod_{\{\rho ^{1}(p,r)\}}^{N_{1}(T)}L(p,r)^{-2H_{j_{(r,p)}}},  \label{scale}
\end{equation}
where the exponents $\{H_{j_{(r,p)}}\}$ characterize the conformal dimension
of the boundary insertion operators $\{\psi _{j_{(r,p)}}^{j_{p}j_{r}}\}$. 
A dynamical triangulation prescription (\emph{i.e.}, holding fixed the $
\{l(p,r)\}$ and simply summing over all possible topological ribbon graphs $
\{\Gamma \}$) feels such a scaling more directly via the two-point function 
(\ref{twopoints}), and (\ref{critedge})(again to be summed over all possible isospin 
channels $j(r,p)$) which exhibit the same exponent dependence.\ 
Even if of great conceptual interest (for a non-critical string view-point), we do not pursue such an
analysis here. We are more interested in discussing, at least at a
preliminary level, how (\ref{FinPart}) relates with the bulk dynamics in the
double $\widetilde{V}_{M}$ of the 3-manifold $V_{M}$ associated with the
triangulated surface $M$. Since we are in a discretized setting, such a
connection manifests itself, not surprisingly, with an underyling structure
of $\ Z^{WZW}(|P_{T_{l}}|,\{\widehat{h}(S_{\theta (i)}^{(+)})\})$ which
directly calls into play, via the presence of the (quantum) $6j$-symbols,
the building blocks of the Turaev-Viro construction. This latter theory is
an example of topological, or more properly, of a cohomological model. When
there are no boundaries, it is characterized by a small (finite dimensional)
Hilbert space of states; in the presence of boundaries, however, cohomology
increases and the model provides an instance of a holographic correspondence
where the space of conformal blocks of the boundary theory (\emph{i.e.}, the
space of pre-correlators of the associated CFT) can be also understood as
the space of physical states of the bulk topological field theory. A
boundary on a Riemann surface, for instance, makes the cohomology bigger and
this is precisely the case we are dealing with since we are representing a
(random Regge) triangulated surface $|T_{l}|\rightarrow M$ \ by means of a
Riemann surface with cylindrical ends. Thus, we come to a full circle: the
boundary discretized degrees of freedom of the $SU(2)$ WZW theory coupled
with the discretized metric geometry of the supporting surface, give rise to
all the elements which characterize the discretized version of the
Chern-Simons bulk theory on $\widetilde{V}_{M}$. What is the origin of such
a Chern-Simons model? The answer lies in the observation that by considering 
$SU(2)$ valued maps on a random Regge polytope, the natural outcome is not
just a WZW model generated according to the above prescription. The
decoration of the pointed Riemann surface $((M;N_{0}),\mathcal{C})$ with the
quadratic differential $\phi $, naturally couples the model with a gauge
field $A$. In order to see explicitly how this coupling works, we observe
that on the Riemann surface with cylindrical ends $\partial M$, associated
with the Regge polytope $|P_{T_{l}}|\rightarrow {M}$, we can introduce $
\mathfrak{su}(2)$ valued flat gauge potentials $A_{(i)}$ locally defined by 
\begin{gather}
A_{(i)}\doteq \gamma _{i}\left[ \sqrt{\phi (i)}\left( \frac{\lambda (i)}{
\kappa }\mathbf{\sigma }_{3}\right) -\frac{\sqrt{-1}}{2\pi }L(i)\left( \frac{
\lambda (i)}{\kappa }\mathbf{\sigma }_{3}\right) d\ln \left| \zeta
(i)\right| \right] \gamma _{i}^{-1}=  \notag \\
\\
=\frac{\sqrt{-1}}{4\pi }L(i)\gamma _{i}\left( \frac{\lambda (i)}{\kappa }
\mathbf{\ \sigma }_{3}\right) \gamma _{i}^{-1}\left( \frac{d\zeta (i)}{\zeta
(i)}-\frac{d\overline{\zeta }(i)}{\overline{\zeta }(i)}\right) ,  \notag
\end{gather}
around each cylindrical end $\Delta _{\theta (i)}^{\ast }$ of base
circumference $L(i)$, and where $\gamma _{i}\in SU(2)$. (It is worthwhile to
note that the geometrical role of $\ $\ the connection $\{A_{(i)}\}$ is more
properly seen as the introduction, on the cohomology group $H^{1}((M,N_{0});
\mathcal{C})$ of the pointed Riemann surface $((M,N_{0});\mathcal{C})$, of
an Hodge structure analogous to the classical Hodge decomposition of $
H^{h}(M;\mathcal{C})$ generated by the spaces $\mathcal{H}^{r,h-r}$ of
harmonic $h$-forms on $(M;\mathcal{C})$ of type $(r,h-r)$. Such a
decomposition does not hold, as it stands, for punctured surfaces since $
H^{1}((M,N_{0});\mathcal{C})$ can be odd-dimensional, but it can be replaced
by the mixed Deligne-Hodge decomposition). The action $S_{|T_{l=a}|}^{WZW}(
\eta )$ gets correspondingly dressed according to a standard prescription
(see \emph{e.g.} \cite{gawedzki}) and one is rather naturally led to the
familiar correspondence between states of the bulk Chern-Simons theory
associated with the gauge field $A$, and the correlators of the boundary WZW
model.


 Let us also stress that the relation between (\ref{FinPart}) and a triangulation of the
bulk 3-manifold $\widetilde{V}_{M}$, say,  the association of tetrahedra to
the (quantum) $6j$-symbols characterized by (\ref{seigei}), is rather
natural under the doubling procedure giving rise to $\widetilde{V}_{M}$ and
to the Schottky double $M^{D}$. Under such doubling, the trivalent vertices $
\{\rho ^{0}(p,q,r)\}$ of \ $|P_{T_{l}}|\rightarrow {M}$ yield two preimages
in $\widetilde{V}_{M}$, say $\sigma _{(3)}^{0}(\alpha )$ and $\sigma
_{(3)}^{0}(\beta )$, whereas the outer boundaries $S_{\theta (p)}^{(+)}$, $
S_{\theta (q)}^{(+)}$, $S_{\theta (r)}^{(+)}$ associated with the vertices $
\sigma ^{0}(p)$, $\sigma ^{0}(q)$, and $\sigma ^{0}(r)$ in $
|T_{l}|\rightarrow M$ are left fixed under the involution $\Upsilon $
defining $M^{D}$. Fix our attention on $\sigma _{(3)}^{0}(\alpha )$, and let
us consider the tetrahedron $\sigma _{(3)}^{3}(p,q,r,\alpha )$ with base the
triangle $\sigma ^{2}(p,q,r)\in |T_{l}|\rightarrow M$ and apex $\sigma
_{(3)}^{0}(\alpha )$. According to our analysis of the insertion operators $
\{\psi _{j_{(r,p)}}^{j_{p}j_{r}}\}$, to the edges $\sigma ^{1}(p,q)$, $
\sigma ^{1}(q,r)$, and $\sigma ^{1}(r,p)$ of the triangle  
$\sigma ^{2}(p,q,r)$ we must
associate the primary labels $j(p,q)$, $j(q,r)$, and $j(r,p)$, respectively.
Similarly, it is also natural to associate with the edges $\sigma
_{(3)}^{1}(p,\alpha )$, $\sigma _{(3)}^{1}(q,\alpha )$, and $\sigma
_{(3)}^{1}(r,\alpha )$ the labels $j_{p}$, $j_{q}$, and $j_{r}$,
respectively. Thus, we have the tetrahedron labelling 
\begin{equation}
\sigma _{(3)}^{3}(p,q,r,\alpha )\longmapsto \left(
j(p,q),j(q,r),j(r,p);j_{p},j_{q},j_{r}\right) .
\end{equation}
The standard prescription for associating the (quantum) $6j$-symbols to a $
SU(2)_{Q}$-labelled tetrahedron such as $\sigma _{(3)}^{3}(p,q,r,\alpha )$
provides 
\begin{equation}
\sigma _{(3)}^{3}(p,q,r,\alpha )\longmapsto \left\{ 
\begin{array}{ccc}
j_{(q,p)} & j_{p} & j_{q} \\ 
j_{r} & j_{(q,r)} & j_{(p,r)}
\end{array}
\right\} _{Q=e^{\frac{\pi }{3}\sqrt{-1}}},
\end{equation}
which (up to symmetries) can be identified with (\ref{seigei}). \ In this
connection, one can observe that the partition function (\ref{FinPart}) has
a formal structure not too dissimilar (in its general representation
theoretic features) from the boundary partition function discussed in 
\cite{arcioni}, but we postpone to a forthcoming paper a detailed analysis of such a
correspondence since it needs to be framed within the broader context of a
study of the properties of the Chern-Simons bulk states associated to (\ref
{FinPart}).

\bigskip

\noindent\textbf{Acknowledgements}

\vspace{0.1cm}

This work was supported in part by the Ministero dell'Universita' e della
Ricerca Scientifica under the PRIN project \emph{The geometry of integrable
systems.} The work of G. Arcioni is supported in part by the European
Community's Human Potential Programme under contract HPRN-CT-2000-00131
Quantum Spacetime. M. Carfora is grateful to P. Di Francesco and N. Kawamoto for constructive remarks at early stages of this work.


\bigskip

\thebibliography{}

\bibitem{'thooft}
G.~'t Hooft,
\emph{``The scattering matrix approach for the quantum black hole: An overview,''}
Int.\ J.\ Mod.\ Phys.\ A {\bf 11} (1996) 4623
[arXiv:gr-qc/9607022].

\bibitem{'thooft2}
G.~'t Hooft,
\emph{``TransPlanckian particles and the quantization of time,''}
Class.\ Quant.\ Grav.\  {\bf 16} (1999) 395
[arXiv:gr-qc/9805079].

\bibitem{'thooft3}
G.~'t Hooft,
\emph{``Quantum gravity as a dissipative deterministic system,''}
Class.\ Quant.\ Grav.\  {\bf 16} (1999) 3263
[arXiv:gr-qc/9903084].

\bibitem{regge}
T.~Regge and R.~M.~Williams,
\emph{``Discrete Structures In Gravity,''}
J.\ Math.\ Phys.\  {\bf 41} (2000) 3964
[arXiv:gr-qc/0012035].

\bibitem{arcioni} G.~Arcioni, M.~Carfora, A.~Marzuoli and M.~O'Loughlin,
\emph{``Implementing holographic projections in Ponzano-Regge gravity,''}
Nucl.\ Phys.\ B {\bf 619} (2001) 690
[arXiv:hep-th/0107112].

\bibitem{various}
L.~Freidel and K.~Krasnov,
\emph{``2D conformal field theories and holography,''}
[arXiv:hep-th/0205091],
M.~O'Loughlin,
\emph{``Boundary actions in Ponzano-Regge discretization, quantum groups and  AdS(3),''}
[arXiv:gr-qc/0002092],
D.~Oriti,
\emph{``Boundary terms in the Barrett-Crane spin foam model and consistent  gluing,''}
Phys.\ Lett.\ B {\bf 532} (2002) 363
[arXiv:gr-qc/0201077].

\bibitem{kawamoto} N. Kawamoto, H. B. Nielsen, N. Sato, \emph{``Lattice Chern-Simons gravity via Ponzano-Regge model''},
Nuc. Phys. B {\bf 555} (1999) 629.
 
\bibitem{carfora} M. Carfora, A. Marzuoli, \emph{``Conformal modes in simplicial quantum
gravity and the Weil-Petersson volume of moduli space''}, [arXiv:math-ph/0107028] 
Adv.Math.Theor.Phys. {\bf 6}(2002) 357.

\bibitem{carfora2} M. Carfora, C. Dappiaggi, A. Marzuoli,  \emph{``The modular geometry of
random Regge triangulations''}, [arXiv:gr-qc/0206077] Class. Quant. Grav. {\bf 19}(2002) 5195.

\bibitem{carfora3} M. Carfora, \emph{``Discretized Gravity and the SU(2) WZW model''}, to appear in Class. Quant. Grav. (2003).

\bibitem{gawedzki} K. Gaw\c edzki, \emph{``Conformal field theory: a case study''}. In: \emph{''Conformal Field Theory''}, 
Frontiers in Physics {\bf 102}, eds. Nutku, Y.,
Saclioglu, C., Turgut, T., Perseus Publ., Cambridge Ma. (2000), 1-55.

\bibitem{gaberdiel} M.R. Gaberdiel, A. Recknagel, G.M.T. Watts, 
\emph{``The conformal boundary states for SU(2) at level 1''}, Nuc. Phys. B {\bf 626} (2002) 344 [hep-th/0108102].


\bibitem{lewellen} D.C. Lewellen, \emph{``Sewing constraints for conformal field
theories on surfaces with boundaries'',} 
Nuc. Phys. B {\bf 372} (1992) 654.

\bibitem{sagnotti} G. Pradisi, A. Sagnotti, Ya. S. Stanev, 
\emph{``Completeness conditions for boundary operators in 2D conformal field
theory'',} 
Phys. Lett. {\bf B 381} (1996) 97.

\bibitem{felder} 
G.~Felder, J.~Frohlich, J.~Fuchs and C.~Schweigert,
\emph{``The geometry of WZW branes,''}
J.\ Geom.\ Phys.\  {\bf 34} (2000) 162
[arXiv:hep-th/9909030].

\bibitem{gaume} L. Alvarez-Gaum\'{e}, C. Gomez, G. Sierra, 
\emph{``Quantum group interpretation of some conformal field theories''}, Phys. Lett. B 220 (1989) 142.

\bibitem{ambjorn} J. Ambj\"orn, B. Durhuus, T. Jonsson, \emph{``Quantum Geometry''},
Cambridge Monograph on \ Mathematical Physics, Cambridge Univ. Press
(1997).

\bibitem{mulase} M. Mulase, M. Penkava, \emph{``Ribbon graphs, quadratic differentials on
Riemann surfaces, and algebraic curves defined over''}$\overline{\mathbb{Q}}$,
The Asian Journal of Mathematics {\bf 2}, 875-920 (1998)  [math-ph/9811024 v2].

\bibitem{Thurston} W. P. Thurston, \emph{``Three-Dimensional Geometry and Topology''} ,(ed. By S. Levy), Princeton Math. Series, Princeton Univ. Press, Princeton, New Jersey (1997).

\bibitem{witten}
E.~Witten,
\emph{``Nonabelian Bosonization In Two Dimensions,''}
Commun.\ Math.\ Phys.\  {\bf 92} (1984) 455.

\bibitem{alekseev}
A. Yu. Alekseev, V. Schomerus, 
\emph{``D-branes in the WZW model'',}
Phys.Rev. D{\bf 60} 061901 hep-th/9812193.

\bibitem{gawedzki2}
K.~Gawedzki,
\emph{``Boundary WZW, G/H, G/G and CS theories,''}
Annales Henri Poincare {\bf 3} (2002) 847
[arXiv:hep-th/0108044].

\bibitem{schomerus} V.~Schomerus,
\emph{``Lectures on branes in curved backgrounds,''}
Class.\ Quant.\ Grav.\  {\bf 19} (2002) 5781
[arXiv:hep-th/0209241].

\bibitem{gaberdiel2}
M.~R.~Gaberdiel,
\emph{``D-Branes From Conformal Field Theory,''}
Fortsch.\ Phys.\  {\bf 50} (2002) 783.
\end{document}